\documentclass[aps,preprintnumbers,showpacs,amsmath,amssymb,floatfix,10pt,prd,superscriptaddress,nofootinbib,twocolumn]{revtex4-2}
\usepackage[utf8]{inputenc}
\usepackage{float}
\usepackage{hyperref}
\hypersetup{
	colorlinks=true,
	linkcolor=blue,
	filecolor=magenta,
	urlcolor=magenta,
}
\usepackage{tabularx}  
 \usepackage{booktabs}
\usepackage{mathtools}
\usepackage{amssymb}
\usepackage{amsmath}
\usepackage{graphicx}
\usepackage{latexsym}
\usepackage{orcidlink}

\begin{document}

\title{ViT-NeBLa: A Hybrid Vision Transformer and Neural Beer–Lambert Framework for Single-View 3D Reconstruction of Oral Anatomy from Panoramic Radiographs}

\author{Bikram Keshari Parida \orcidlink{0000-0003-1204-357X}}
\email{parida.bikram90.bkp@gmail.com}
\affiliation{Artificial Intelligence and Image Processing Lab., Department of Information and Communication Engineering, Sun Moon University, South Korea}

\author{Anusree P. Sunilkumar \orcidlink{0009-0006-9381-3618}}
\email{anusreepandath@gmail.com}
\affiliation{Artificial Intelligence and Image Processing Lab., Department of Information and Communication Engineering, Sun Moon University, South Korea}

\author{Abhijit Sen \orcidlink{0000-0003-2783-1763}}
\email{asen1@tulane.edu}
\affiliation{Department of Physics and Engineering Physics, Tulane University, New Orleans, LA 70118, USA}

\author{Wonsang You \orcidlink{0000-0002-6806-7135}}
\email{wyou@kaist.ac.kr}
\affiliation{Artificial Intelligence and Image Processing Lab., Department of Information and Communication Engineering, Sun Moon University, South Korea}

\begin{abstract}
	
Dental diagnosis relies on two primary imaging modalities: panoramic radiographs (PX) providing 2D oral cavity representations, and Cone-Beam Computed Tomography (CBCT) offering detailed 3D anatomical information. While PX images are cost-effective and accessible, their lack of depth information limits diagnostic accuracy. CBCT addresses this but presents drawbacks including higher costs, increased radiation exposure, and limited accessibility. Existing reconstruction models further complicate the process by requiring CBCT flattening or prior dental arch information, often unavailable clinically. We introduce ViT-NeBLa, a vision transformer-based Neural Beer-Lambert model enabling accurate 3D reconstruction directly from single PX. Our key innovations include: (1) enhancing the NeBLa framework with Vision Transformers for improved reconstruction capabilities without requiring CBCT flattening or prior dental arch information, (2) implementing a novel horseshoe-shaped point sampling strategy with non-intersecting rays that eliminates intermediate density aggregation required by existing models due to intersecting rays, reducing sampling point computations by $52 \%$, (3) replacing CNN-based U-Net with a hybrid ViT-CNN architecture for superior global and local feature extraction, and (4) implementing learnable hash positional encoding for better higher-dimensional representation of 3D sample points compared to existing Fourier-based dense positional encoding. Experiments demonstrate that ViT-NeBLa significantly outperforms prior state-of-the-art methods both quantitatively and qualitatively, offering a cost-effective, radiation-efficient alternative for enhanced dental diagnostics.\\



\end{abstract}

\maketitle

\section{Introduction}


The intricate anatomical details of the oral cavity are paramount for accurate dental diagnostics and effective treatment planning. While traditional 2D imaging techniques, such as the panoramic X-ray (PX), have long served as essential tools for initial assessment, they inherently lack the depth information crucial for comprehensive understanding \cite{scarfe2006,angelopoulos_2012}. The transition towards 3D representations in dental imaging offers a significant leap forward, enabling clinicians to visualize the complex spatial relationships between teeth, bone, and surrounding soft tissues with greater clarity \cite{jacobs_cone_2018}. This enhanced visualization capability has profound implications for various clinical applications, including the precise planning of dental implant surgeries, the accurate assessment of impacted teeth requiring extraction, and the detailed evaluation necessary for orthodontic treatments \cite{ganz_cone_2011}. Furthermore, 3D models serve as invaluable tools for patient education, facilitating a better understanding of their condition and the proposed treatment strategies \cite{liang_2021a}. This shift from a flattened view to a volumetric understanding mirrors a broader trend in medical imaging, where the richness of 3D data provides far more insightful information for diagnosis and intervention \cite{scarfe2006}.

CBCT  has emerged as the gold standard in dental imaging for acquiring detailed 3D volumetric data of the oral and maxillofacial region \cite{scarfe2006}. CBCT provides high-resolution images of hard tissues, allowing for precise measurements and comprehensive anatomical evaluation \cite{gupta_cone_2013}. However, despite its undeniable advantages, the routine use of CBCT is often limited by its higher cost and the increased radiation dose compared to conventional 2D radiographs \cite{ludlow_2015,oenning_2018}. This is a critical consideration, as the principle of ALARA (As Low As Reasonably Achievable) dictates that medical imaging should employ the lowest possible radiation dose while still providing adequate diagnostic information. Consequently, there exists a significant need to explore alternative methodologies that can offer valuable 3D insights with lower radiation exposure and cost \cite{park_2024}.

PX, also known as panoramic radiographs, are widely employed in dental practice due to their broad availability, cost-effectiveness, and the substantially lower radiation dose they deliver compared to CBCT \cite{vandenberghe_2010}. A PX captures the entire maxillofacial region, including all the teeth in both the upper and lower jaws, as well as the surrounding bone structures, in a single image \cite{shah_recent_2014}. This comprehensive view makes it a common and efficient initial diagnostic tool for a wide range of dental conditions. If it were possible to accurately reconstruct 3D oral structures from a single PX, it would leverage the inherent benefits of PX while simultaneously overcoming its limitations in providing depth information, thus significantly expanding its clinical utility \cite{park_2024,s_deep_2023}. Achieving such a capability could democratize access to valuable 3D information in dental care, particularly in regions or settings where CBCT technology is not readily accessible or financially feasible.

However, the task of inferring a 3D structure from a single 2D projection, such as a PX, presents a fundamentally challenging and ill-posed problem \cite{s_deep_2023, x2teeth}. The process of projection inherently leads to a loss of information, specifically the depth information along the direction of the X-ray beam. Furthermore, the significant anatomical variations observed across different patients add another layer of complexity to the reconstruction task \cite{vandenberghe_2010}. A single 2D view might represent a multitude of possible 3D configurations, making it difficult to uniquely determine the underlying structure \cite{park_2024}. PXs themselves introduce geometric distortions due to the specific trajectory of the moving X-ray source and detector, which follows the curvature of the dental arch \cite{stramotas_accuracy_2002}. This complex acquisition process makes direct geometric inversion to obtain a 3D model exceptionally challenging \cite{gribel_accuracy_2011}.  Finally, the reconstruction of fine details, such as the intricate root morphology of teeth and the precise boundaries of various oral structures, from a flattened 2D image poses a considerable hurdle. The inherent ambiguity in this single-view 3D reconstruction necessitates the integration of prior knowledge about the typical shapes and arrangements of oral structures, or the development of sophisticated learning models capable of inferring the missing 3D information from the 2D input \cite{park_2024,x2teeth}.

The advent of deep learning has brought about a paradigm shift in the field of medical image analysis, demonstrating remarkable success in a wide array of tasks, including the challenging problem of image reconstruction \cite{litjens_survey_2017,shen_deep_2017,bai_2024}. Convolutional Neural Networks (CNNs), a specific type of deep learning architecture, have proven particularly effective in learning complex features directly from medical images and have shown considerable promise in the task of 3D reconstruction from limited views, including single 2D images \cite{esteva_guide_2019,chen_deep_2020,song_2021,liang_2021a}. These deep learning models possess the capability to learn the intricate and often non-linear mapping that exists between 2D projections and the underlying 3D structures by being trained on large datasets consisting of paired 2D and 3D images \cite{yaqub_deep_2022}. This data-driven approach allows the models to implicitly learn the statistical properties of oral structures and how they are projected in PXs \cite{naf_2022,song_2023}.

The application of deep learning to dental reconstruction has evolved significantly over time. Initial efforts focused on more fundamental tasks, such as the accurate segmentation and identification of individual teeth within PXs \cite{jader_deep_2018,lee_diagnosis_2018}. These studies demonstrated the high accuracy that deep learning models could achieve in these tasks, paving the way for more complex applications. Subsequent research endeavors explored the possibility of reconstructing 3D representations of either individual teeth or the entire oral cavity directly from PXs using various CNN-based architectures \cite{yaqub_deep_2022,s_deep_2023,liang_2021a}. For instance, the X2Teeth model utilized a CNN-based architecture to decompose the complex reconstruction task into two sub-problems: first, the localization of individual teeth within the panoramic image, and second, the estimation of the 3D shape of each localized tooth \cite{x2teeth}. This approach yielded promising quantitative results, demonstrating the potential of CNNs for this challenging task. Another notable contribution is the Oral-3D framework, which employed Generative Adversarial Networks (GANs) to learn the transformation from a 2D PX into a flattened 3D representation of the oral cavity \cite{song_2021}. This was followed by a deformation module that utilized prior information about the typical shape of the dental arch to restore the natural curvature of the mandible in the final 3D reconstruction. The evolution of deep learning in dental reconstruction clearly shows a progression from tackling basic image analysis tasks to addressing the more intricate challenge of holistic 3D reconstruction, with increasing sophistication in the design of model architectures and the development of novel methodologies.

More recently, Vision Transformers (ViTs) have emerged as a powerful and highly competitive architecture in the broader field of computer vision \cite{dosovitskiy_image_2021}. These models have demonstrated state-of-the-art (SOTA) performance across a wide range of tasks, including image recognition, object detection, and image synthesis \cite{khan_transformers_2022}. A key advantage of ViTs lies in their ability to capture long-range dependencies within images through the use of a self-attention mechanism \cite{liu_swin_2021,vit_v_net,vit_nerf}. This allows them to understand the global context of an image more effectively compared to traditional CNNs, which primarily focus on local features. In parallel, Neural Radiance Fields (NeRF) have been introduced as a novel approach for 3D scene modeling \cite{mildenhall_nerf_2022}. NeRF works by implicitly learning the radiance field of a scene, which is a continuous volumetric representation that describes the light emitted from every point in space. This learned radiance field can then be used to render photorealistic images of the scene from arbitrary viewpoints \cite{park_2024}. NeRF and similar implicit neural representation methods have shown significant potential for 3D reconstruction from limited sets of 2D views in medical imaging, offering the capability to generate detailed and continuous representations of anatomical structures \cite{nerf_survey,NeAS,ToothInpaintor,naf_2022,greenspan_learning_2023,molaei_implicit_2023,snaf,song_2023,Park_2023a}. The recent successes of ViTs in understanding global image context and NeRF in generating high-fidelity 3D representations suggest their potential to address some of the inherent limitations of previous deep learning models when applied to the problem of 3D reconstruction from a single PX \cite{vit_nerf,park_2024,song_2023}.

Motivated by the complementary strengths of ViTs for robust feature extraction and NeRF-like implicit representations for high-quality 3D reconstruction, we propose a novel deep learning model named ViT-NeBLa. Our model is specifically designed to leverage the global context understanding capabilities of ViTs \cite{dosovitskiy_image_2021,vit_nerf} to guide an implicit neural representation-based reconstruction process \cite{park_2024}. The goal is to achieve accurate 3D estimation of oral structures from a single PX radiograph, without relying on CBCT flattening and supplementary data like dental arch curves. We hypothesize that the synergistic combination of the ViT's attention mechanism, which allows the model to focus on relevant features across the entire panoramic image, and NeBLa's continuous volumetric representation, which enables the generation of detailed and accurate 3D geometry, will lead to significant improvements in reconstruction accuracy and level of detail compared to existing SOTA methods. The motivation behind ViT based NeBLa is to effectively combine the advantages of transformer-based architectures in capturing global context with the detailed reconstruction capabilities offered by implicit neural representations, thereby advancing the current SOTA in 3D dental reconstruction from widely available single PXs \cite{3dentai,3dpx_2024}.

\textbf{Contributions of this work:} In our model, we have incorporated a ViT-based feature extraction module that effectively extracts global features. Additionally, we have included a CNN-based module that specializes in local feature extraction. Moreover, we have implemented learnable positional hash encoding to represent positional information in higher dimensions.  Another main difference between other existing SOTA papers and ours is that we employed an elliptical trajectory, wherein the rays were deemed to be tangent to this particular path, thereby removing the necessity for additional data like dental arch curves. The pixel sampling conducted along these rays was strictly restricted to a horseshoe-shaped focal region that effectively encompasses the jawline. This strategic confinement significantly limits the number of sample points per ray, which in our specific case is $96$, distinctly 52 \% less than the $200$ sample points used in the NeBLa paper \cite{park_2024}. This approach ensures a balance between precision and computational efficiency in our study.

The rest of the paper is organized as follows: Section ~\ref{sec:rel_work} provides a detailed review of the existing research works related to 3D dental reconstruction from single PXs using deep learning. Section ~\ref{sec:method} elaborates on the architecture and methodology of the proposed ViT based NeBLa model. Section ~\ref{sec:expt} presents the experimental setup and the results obtained. Section ~\ref{sec:l&fw} discusses the limitations and future works.  Finally, section ~\ref{sec:con} concludes the paper and outlines potential directions for future research in this domain.



\section{Related Works}\label{sec:rel_work}

Reconstructing 3D oral anatomy from a single PX is inherently challenging, but recent deep learning methods have made significant progress. Liang et al. pioneered full-mouth 3D reconstruction from a single PX with X2Teeth, which detects individual teeth and predicts each tooth’s 3D shape \cite{liang2021}. By breaking the task into single-tooth sub-problems, X2Teeth captured fine dental details and significantly improved accuracy over direct volumetric prediction \cite{liang2021}. However, it reconstructs only teeth and requires paired PX–CBCT data for training.


Subsequent methods introduced anatomical priors to recover more complete structures. Song et al. proposed Oral-3D, a two-stage model guided by a predefined dental arch curve to reconstruct the full oral cavity \cite{song_2021}. The arch prior ensured a plausible jaw shape and a refinement stage enhanced tooth details. Oral-3D achieved better fidelity than baseline CNNs \cite{song_2021}, but it relies on manual arch extraction and can falter on real PX images due to domain differences. Liang et al. extended this line with OralViewer, an interactive system that combines PX-based 3D teeth reconstruction with generic gum and jaw models for patient education \cite{liang_2021a}. OralViewer showed improved tooth reconstruction detail \cite{liang_2021a}, but the 3D registration of the jaw was still performed using predefined dental arch templates rather than patient-specific anatomical data.

To reduce reliance on scarce 3D supervision, recent approaches incorporate X-ray physics and implicit representations. Physics-based methods like NeBLa and Oral-3Dv2 eliminate the need for matched CBCT data by integrating the PX projection model into training \cite{park_2024, song_2023}. These models achieve state-of-the-art accuracy without requiring arch shape priors, albeit with increased training complexity \cite{park_2024, song_2023}. Other innovations include 3DPX, which uses a hybrid MLP–CNN multi-scale pipeline to improve depth reasoning and consistency over CNN-only models \cite{3dpx_2024}. Similarly, 3DentAI embeds PX geometry into its attention U-Net pipeline by first predicting a flattened jaw volume and then warping it to the true curved jaw shape for refinement \cite{3dentai}. Using only synthetic PX from CBCT scans, 3DentAI achieves good results without any real PX–CBCT pairs \cite{3dentai}, though it assumes a normal jaw curvature and complete dentition.

Beyond full reconstructions, other work leverages PX to augment partial 3D data. ToothInpaintor uses a PX image to fill in missing teeth on an existing 3D dental scan, aligning the completed model’s projection with the radiograph \cite{ToothInpaintor}. This illustrates how 2D radiographs can guide completion of partial 3D structures.

The latest, PX2Tooth, takes a tooth-centric approach. It segments each tooth in the PX and then reconstructs that tooth as a 3D point cloud \cite{px2tooth_2024}. With a large training set ($\sim$ 500 paired cases), PX2Tooth set a new benchmark for accuracy \cite{px2tooth_2024}, though it still focuses only on teeth and depends on extensive training data.


In summary, the field has evolved from basic CNN mappings to sophisticated methods that exploit anatomical knowledge, X-ray physics, and improved networks. Each approach tackles key challenges—handling multiple teeth, limited data, or realism—yet limitations remain. Many methods reconstruct only the dentition or require extra inputs (arch curves, templates) or large supervised datasets, and few capture patient-specific bone anatomy. These gaps motivate our work, which aims to reconstruct a comprehensive 3D oral structure (teeth and bone) from a single PX without requiring the intermediate steps like prediction of flattened CBCT volume and without requiring dental arch curve info building on insights from prior studies.

\section{Methodology}\label{sec:method}

\begin{figure*}[htbp!]
	\centering
	\includegraphics[width=\linewidth]{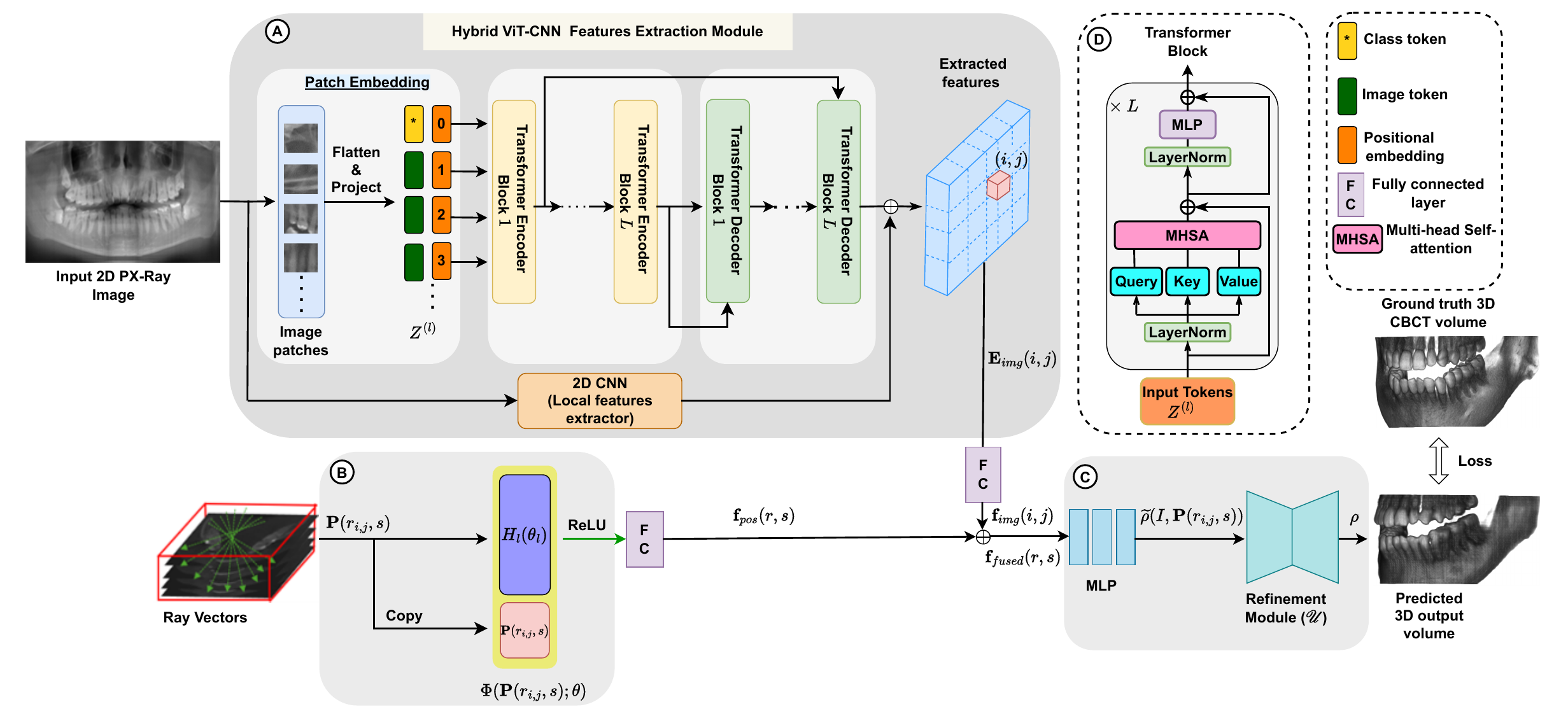}
	\caption{\textbf{Schematic overview of our proposed pipeline.} (A) A hybrid ViT-CNN feature extractor, which divides the input  PX image $I \in \mathbb{R}^{1 \times H \times W}$ into $P \times P$ patches, embeds then with positional and class tokens, and process them through transformer and convolutional branches to extract dense global and local feature maps $\mathbf{E}_{img}(i,j)$. Concurrently, (B) 3D query points $\mathbf{P}(r_{i,j},s) \in \mathbb{R}^{3}$ are sampled along horseshoe-shaped focal rays $r_{i,j}$ tangent to an elliptical trajectory and encoded via a learnable multi-resolution hash grid $\Phi(\mathbf{P}(r_{i,j},s);\theta)$. (C) At each sample location, image features $\mathbf{f}_{img}(i,j)$ and positional encoding $\mathbf{f}_{pos}(r,s)$ are fused and fed into a compact multi-layer perceptron (MLP) to predict volumetric densities $\widetilde{\rho}(r,s)$. Standard volume rendering produces a coarse implicit CBCT volume, which is then refined by a  3D U-Net ($\mathcal{U}$) to yield the final high-fidelity 3D reconstruction of oral anatomy. (D) The transformer block applies $L$ layers of self-and cross-attention (with residual connections and layer-norm) to the class-augmented tokens.}
	\label{fig:framework}
\end{figure*}

In this section, we will discuss the overall implementation of our proposed model for 3D reconstruction of oral anatomoy from single PX input image; and details of the sub-modules (the building blocks of our model) that make up the model.  The pipeline of the proposed framework is outlined in the Fig. \ref{fig:framework}.

\subsection*{Implemented Model Summary}

Our aim is to model single-view 3D reconstruction as a continuous mapping
\begin{align}
    F:\bigl(I,\mathbf{P}\bigr)\longmapsto\rho,
\end{align}

where $I$ is a single 2D PX image and $\mathbf{P}$ a set of 3D query points, yielding a volumetric density field $\rho$. Concretely, let $I\in\mathbb{R}^{C\times H\times W},\quad
C=1 \, \text{channel},\;H=128 \, \text{height},\;\text{and}\, W=256 \, \text{width}$. We define the set of rays stimulated tangentially along a pre-defined elliptical trajectory within the CBCT volume (See Fig. \ref{fig:ray_sampling}(a)), and it is expressed as,
\begin{align}
    \mathcal{R}=\{r_{i,j}| \, i=1,\dots,H,\;j=1,\dots,W\}, \label{set_ray}
\end{align}

so that $  |\mathcal{R}|=H\times W$. Here, $r_{i,j}$ denotes the stimulated ray for the rendered pixel $(i,j)$ in $I$. When we later refer to a generic $r \in \mathcal{R}$, it denotes one of these $r_{i,j}$.  For each ray $r \in \mathcal{R}$, we sample the set of points $\mathcal{S}=\{s_{1},s_{2},\dots,s_{96}\}$, so that $|\mathcal{S}| = 96$. Here, each $s_{k}$ specifies the $k$-th sample location along the ray;  when we write a generic $s \in \mathcal{S}$, it denotes one of these sample indices.  Thus the query tensor is $\mathbf{P}\in\mathbb{R}^{|\mathcal{R}|\times|\mathcal{S}|\times 3},$ collecting coordinates $\mathbf{P}(r,s)\in\mathbb{R}^3$ for every $r \in \mathcal{R}$ and $s \in \mathcal{S}$.  The target volume is $\rho\in\mathbb{R}^{C\times H\times W\times D},\, \text{with depth}\, D=256$.

\subsection*{Implemented Modules Summary}

Before we delve into the details of each modules that make up the entire model, let us quickly summarize the overal idea behind each module. The proposed model consists of four major modules. In Fig. \ref{fig:framework}(A), depicting first module, we  extract a dense image feature map  $\mathbf{E}_{\mathrm{img}} = \mathcal{F}(I)\;\in\;\mathbb{R}^{H\times W\times \kappa},$ where $\mathcal{F}$  is a ViT-based hybrid-features extractor. Each spatial vector $\mathbf{E}_{\mathrm{img}}(i,j)$  is lifted by linear projector $\phi_{\mathrm{img}}\colon\mathbb{R}^{\kappa}\to\mathbb{R}^f$ 
 into a $f$-dimensional embedding $\mathbf{f}_{\mathrm{img}}(i,j)
=\phi_{\mathrm{img}}\bigl(\mathbf{E}_{\mathrm{img}}(i,j)\bigr)
\in\mathbb{R}^f$. In parallel, each 3D coordinate \(\mathbf{P}(r,s)\) is encoded by a learnable multi-resolution hash encoder
\(\Phi\colon\mathbb{R}^3\to\mathbb{R}^{3 + \zeta}\) (Fig. \ref{fig:framework}(B)), then linearly mapped by
\(\phi_{\mathrm{pos}}\colon\mathbb{R}^{3 + \zeta}\to\mathbb{R}^f\) into positional features $\mathbf{f}_{\mathrm{pos}}(r,s)
=\phi_{\mathrm{pos}}\bigl(\Phi(\mathbf{P}(r,s))\bigr)
\in\mathbb{R}^f$.

At each sample point we fuse image and positional features: $\mathbf{f}_{\mathrm{fused}}(r,s)
=\mathbf{f}_{\mathrm{img}}\bigl(i(r),j(r)\bigr)
+\mathbf{f}_{\mathrm{pos}}(r,s),$ where for ray $r=r_{i,j}$  projects to pixel $(i,j)$ in $I$. Finally, a compact MLP $\mathcal{M}\colon\mathbb{R}^f\to\mathbb{R}$ predicts scalar densities
$\rho(r,s)=\mathcal{M}\bigl(\mathbf{f}_{\mathrm{fused}}(r,s)\bigr),$
and standard volume rendering along all rays in $\mathcal{R}$ reconstructs a coarse implicit full 3D field $\widetilde{\rho}$, which is then refined by a 3D U-Net ($\mathcal{U}$) to yield the final high-fidelity 3D volumetric oral anatomy $\rho$. The third and final modules are illustrated in Fig. \ref{fig:framework}(C). By training $\mathcal{F},\Phi,\phi_{\mathrm{img}},\phi_{\mathrm{pos}}$, $\mathcal{M}$ and $\mathcal{U}$ end-to-end, the model learns to lift flat X-ray textures into coherent 3D volumetric oral structure.

 Hereafter, in the subsequent subsections, we will discuss the implementation details of each module which includes mathematical depictions of each module along with their quantitative configurations employed. 
 
\subsection*{Point sampling from Rays}

Before delving into the detailed architecture of the proposed modules, which we defer to the next section, we first outline a critical methodological step undertaken in this work: the generation of synthetic PX images ($I$) and the corresponding 3D sample points along each ray, as defined in Eq.~\eqref{set_ray}. These sample points represent the discrete locations in 3D space that contribute to the formation of the synthetic PX image via forward projection, following Beer–Lambert’s law \cite{Ketcham_2014,Max_1995}. Both the synthetic PX images and the associated 3D sample point sets are jointly used as inputs during the training of our proposed model, serving as a foundation for learning the mapping from 2D projections to 3D anatomical structures.

Our approach employs novel point sampling strategies that differ substantially from existing methodologies in the literature. This implementation achieves significant memory reduction while maintaining reconstruction fidelity. Unlike conventional approaches such as those proposed in Park et al \cite{park_2024} and Song et al \cite{song_2023}, our sampling technique utilizes non-intersecting rays through the implementation of horseshoe-shaped-restricted focal region, thereby eliminating the need for intermediate density aggregation steps. This architectural choice contributes to both the computational efficiency and memory optimization of our method.

\begin{figure}[htbp!]
    \centering
    \includegraphics[width=\linewidth]{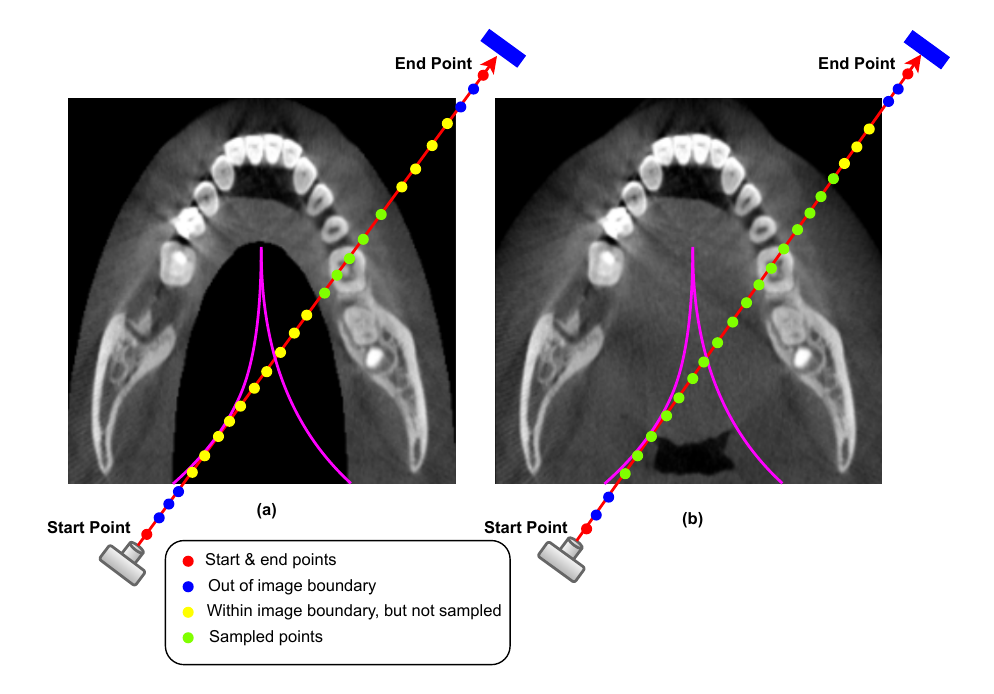}
    \caption{Comparison of ray‐sampling strategies. \textbf{(a)} Proposed method: each ray traverses a horseshoe‐shaped focal region enclosing the jaw, sampling only the green points (96 per ray) while skipping all other in‐boundary locations (yellow). Because these rays remain tangent to the elliptical trajectory (pink) and do not intersect, their sample points never overlap—eliminating the need for an intermediate density aggregation step. \textbf{(b)} NeBLa baseline \cite{park_2024}: each ray samples 200 uniformly spaced points (green) across the full image extent; overlapping sampling requires an explicit intermediate density function to resolve multiple contributions. Both methods use identical inter‐sample spacing, but our focused strategy halves the per‐ray sample count and avoids density‐fusion complexity, significantly reducing computational overhead.}
    \label{fig:ray_sampling}
\end{figure}

Synthetic PX images ($I$) are generated by ray-casting technique through the CBCT volume. For each rendered pixel in $I\in\mathbb{R}^{C\times H\times W} $, a ray is stimulated tangentially along the pre-defined elliptical trajectory as illustrated in Fig. \ref{fig:ray_sampling}(a) (see Appendix \ref{app:dp} for further details). Given the rays $\mathcal{R}$ defined in Eq. ~\eqref{set_ray}, each ray $r_{i,j} \in \mathcal{R}$ is parametrized as:

\begin{align}
	\mathbf{r}_{i,j}(t)
	=\mathbf{o}_{i,j} \;+\; t\,\hat{\mathbf{d}}_{i,j}, 
	\quad t\in[t_{\min},t_{\max}],
\end{align}

where \(\mathbf{o}_{i,j}\) and \(\hat{\mathbf{d}}_{i,j}\) are the origin and unit direction of \(r_{i,j}\).  To concentrate sampling on the mandibular region, we restrict \(t\) to a horseshoe-shaped focal interval $\mathcal{T}\subset[t_{\min},t_{\max}]$  that encloses the jaw. Confining sampling to the anatomically relevant horseshoe-shaped region dramatically reduces sample points versus uniform sampling over entire image volume, lowering memory footprint and enabling faster convergence without sacrificing reconstruction fidelity.

Within this interval $\mathcal{T}\subset[t_{\min},t_{\max}]$, we uniformly sample $S=96$  points using the following equation

\begin{align*}
    t_s = t_{\min} \;+\; \frac{s-1}{S-1}\,\bigl(t_{\max}-t_{\min}\bigr), \quad s=1,\dots,S,
\end{align*}

and collect the query 3D coordinates
\begin{align}
    \mathcal{S}=\{s_1,\dots,s_S\},  \qquad \mathbf{P}(r_{i,j},s) =\mathbf{r}_{i,j}(t_s)\in\mathbb{R}^3. \label{sample_points}
\end{align}

Since $\lvert\mathcal{R}\rvert=H\times W$ and $\lvert\mathcal{S}\rvert=S$, the full tensor of 3D samples has shape $\mathbf{P}\in\mathbb{R}^{|\mathcal{R}|\times|\mathcal{S}|\times3}$.

By stimulating one ray per rendered pixel in PX image $I$ and sampling 96 uniformly spaced points within the anatomically relevant focal window, we obtain a consistent set of 3D queries for subsequent encoding and fusion. Fig. \ref{fig:ray_sampling} highlights the key difference between the voxel sampling strategies of the NeBLa method \cite{park_2024} and our proposed approach. In the left panel, yellow voxels lie within the image boundary but are ignored by our sampling; by contrast, we sample only the green voxels that fall inside the horseshoe-shaped focal region enclosing the jaw. This focused sampling eliminates redundant background points, reducing computational load while preserving anatomically relevant detail.

\subsection*{Detailed Implemented Modules }
\subsection{Position Encoding}
Traditional Fourier‐based positional encodings (e.g.\ sinusoidal embeddings) or dense grid encodings require either large embedding dimensions or prohibitively fine grid resolutions to capture high‐frequency variations, leading to excessive memory overhead and limited scalability \cite{park_2024,bai_2024}. To overcome these limitations, we adopt a different path where we implement learnable multi‐resolution hash encoding that maintains constant memory per level while still representing fine spatial detail \cite{naf_2022}.

Each query coordinate $\mathbf{P}(r_{i,j},s)\in\mathbb{R}^3$, defined in Eq.~\eqref{sample_points}, must be lifted into a compact, multi‐resolution feature space before fusion with image features.  Rather than storing a dense 3D grid, we employ a learnable hash‐grid that balances expressivity and memory efficiency. At level $l=0,\dots,\xi-1$, we first determine a spatial resolution  

\begin{align}
    r_l = \min\bigl(r_0\,2^l,\;r_{\max}\bigr),
\end{align}

where $r_0=16$ is the coarsest grid and $r_{\max}=256$ the finest.  Intuitively, $r_l$ partitions continuous space into cubes of side length $1/r_l$, capturing features at progressively finer scales.
We then discretize the continuous coordinate by  

\begin{align}
    \bigl\lfloor \mathbf{P}(r_{i,j},s)\,r_l\bigr\rfloor
=\bigl(\lfloor x\cdot r_l\rfloor,\lfloor y\cdot r_l\rfloor,\lfloor z\cdot r_l\rfloor\bigr)\in\mathbb{Z}^3, 
\end{align}

mapping $\mathbf{P}(r_{i,j},s)$  into an integer grid index.  This “floor” operation $\lfloor \, \cdot \, \rfloor$ ensures that nearby points fall into the same or adjacent bins, preserving spatial locality.

To embed these indices into a fixed‐size table, we compute a hash index as follows,
\begin{align}
 &h_l\bigl(\mathbf{P}(r_{i,j},s)\bigr) = \nonumber\\
 & \bigl(\lfloor x\cdot r_l\rfloor\, \cdot p_1
\;\oplus\;\lfloor y\cdot r_l\rfloor\, \cdot p_2
\;\oplus\;\lfloor z\cdot r_l\rfloor\, \cdot p_3\bigr)\;\bmod\; S, \label{hash_index}
\end{align}

where $p_1=1$, $p_2=2654435761$, $p_3=805459861$ are large, distinct primes chosen to minimize collisions, $\oplus $ denotes bit‐wise XOR, and modulo  $S=2^{19}$ fixes the table length.  This hashing step spreads spatial indices uniformly across $S$ slots, bounding memory usage regardless of $r_l$.

At each level $l$, a learnable table  $H_l(\theta_l)\in\mathbb{R}^{S\times F}$ stores $F$ -dimensional feature vectors parametrized by $\theta_l$.  After computing the hash index $h_l\bigl(\mathbf{P}(r_{i,j},s)\bigr)$ defined in Eq. \eqref{hash_index}, we retrieve the corresponding feature vector,

\begin{align}
    \mathbf{f}_l\bigl(\mathbf{P}(r_{i,j},s);\theta_l\bigr)
=H_l(\theta_l)\bigl[h_l(\mathbf{P}(r_{i,j},s))\bigr]\;\in\;\mathbb{R}^F.
\end{align}

Each retrieved \(\mathbf{f}_l\) captures local geometry at scale \(l\); as \(l\) increases, the grid refines, encoding finer details. We then concatenate these $\xi$  feature vectors with the original coordinate to get the final encoding vector, representing $\mathbf{P}(r_{i,j},s)$ across scales, is given as
\begin{align}
    \Phi\bigl(\mathbf{P}(r_{i,j},s);\theta\bigr)
=\bigl[\mathbf{P}(r_{i,j},s)\,,\,\mathbf{f}_0,\dots,\mathbf{f}_{L-1}\bigr]
\in\mathbb{R}^{3 + \zeta},
\end{align}

where  $\theta=\{\theta_l\}_{l=0}^{L-1}$ and $\zeta =  \xi \cdot F$.  Finally, a learned linear map   $\phi_{\mathrm{pos}}\colon\mathbb{R}^{3+\zeta}\to\mathbb{R}^f$
 projects this vector into the common feature dimension $f$, yielding $\mathbf{f}_{\mathrm{pos}}(r_{i,j},s)$.

During training, the hash tables adapt via backpropagation. Each $H_l(\theta_l)$  specializes to spatial regions that demand higher fidelity.  Because each lookup and projection costs $\mathcal{O}(1)$  per level, the total encoding complexity is $\mathcal{O}(L)$ per point, and memory per level remains fixed at $S\times F$.  This design ensures a scalable encoding that captures both coarse and fine geometry without exploding memory or compute requirements.  The detailed configurations of this learnable multi-resolution hash positional encoding are given in the table \ref{tab:hash_encoding_config}.

\subsection{Features extraction module}

To produce the per‐pixel feature map \(\mathbf{E}_{\mathrm{img}}=\mathcal{F}(I)\in\mathbb{R}^{H\times W\times \kappa}\), our feature extractor \(\mathcal{F}\) integrates information from both global and local representations by combining three components: (a) a Vision Transformer  branch for capturing global context, (b) a convolutional branch for local detail, and (c) a fusion‐and‐mapping step that aligns both feature types into a unified embedding space \(\mathbb{R}^f\).  This hybrid design is motivated by the complementary strengths of the two architectures. ViTs are particularly effective at modeling long-range dependencies through self-attention mechanisms \cite{liu_swin_2021,vit_v_net,vit_nerf}, enabling a more holistic understanding of the image’s global structure. In contrast, CNNs excel at capturing local spatial patterns through hierarchical receptive fields. By fusing the outputs of both branches, our extractor $\mathcal{F}$ generates rich and spatially coherent representations that leverage the global context from ViTs and the precise local details from CNNs—essential for accurate 3D anatomical reconstruction from single 2D projections.  Our feature extraction module differs from existing approaches: U-Net-based modules efficiently extract local features \cite{zhou_u-net_2020}, while our hybrid ViT-CNN  modules excel at both global and local feature extraction.

\paragraph{\textbf{Global context via ViT encoder–decoder.}}  
We begin by partitioning the input image \(I\in\mathbb{R}^{C\times H\times W}\) (\(C=1\)) into \(N=\tfrac{H\,W}{P^2}\) non‐overlapping patches \(\{x_i\}_{i=1}^N\), each of size \(P\times P\).  Each patch \(x_i\in\mathbb{R}^{P^2C}\) is linearly embedded into \(\mathbb{R}^d\) and augmented with a learned positional vector \(p_i\in\mathbb{R}^d\) as follows:
\[
z_i^{(0)} = W_{\mathrm{emb}}\,x_i + p_i.
\]
We also introduce a learnable class token $z_{cls} \in \mathbb{R}^d$ to capture global information. The initial token matrix then becomes,
\begin{equation}\label{eq:vit-init}
Z^{(0)} = \bigl[z_{\mathrm{cls}}, z_1^{(0)},\dots,z_N^{(0)}\bigr]\;\in\;\mathbb{R}^{(N+1)\times d}.
\end{equation}
Here \(W_{\mathrm{emb}}\in\mathbb{R}^{d\times P^2C}\) projects each flattened patch into the \(d\)-dimensional embedding space, and the class token sits in the first row of $Z^{(0)}$.  The concatenation in Eq.~\eqref{eq:vit-init} collects all patch embeddings into a single matrix for subsequent Transformer processing.

An \(L\)-layer encoder then alternates multi‐head self‐attention (MHSA) and feed‐forward (MLP) blocks, each wrapped in residual connections and layer‐norm (LN) as follows:
\[
\begin{aligned}
\widetilde{Z}^{(\ell)} &= \mathrm{MHSA}\bigl(\mathrm{LN}(Z^{(\ell-1)})\bigr) + Z^{(\ell-1)},\\
Z^{(\ell)} &= \mathrm{MLP}\bigl(\mathrm{LN}(\widetilde{Z}^{(\ell)})\bigr) + \widetilde{Z}^{(\ell)},
\end{aligned}
\quad \ell=1,\dots,L.
\]
We retain each encoder output \(E^{(\ell)}=Z^{(\ell)}\) for skip connections in the decoder. The transformer block is illustrated in Fig. \ref{fig:framework}(D).

The decoder mirrors this structure with \(L\) cross‐attention layers, each attending to its corresponding encoder state. Mathematically expressed as,

\begin{align*}
\widetilde{D}^{(\ell)} = \mathrm{MHA}\bigl(\mathrm{LN}(D^{(\ell-1)}),\,K=E^{(L-\ell+1)},\,V=E^{(L-\ell+1)}\bigr)\\
+ D^{(\ell-1)},
\end{align*}
\begin{align*}
    D^{(\ell)} &= \mathrm{MLP}\bigl(\mathrm{LN}(\widetilde{D}^{(\ell)})\bigr) + \widetilde{D}^{(\ell)},
\quad \ell=1,\dots,L,
\end{align*}

with \(D^{(0)}=Z^{(L)}\).  Following the last cross‐attention layer, we take the token embeddings \(D^{(L)}\in\mathbb{R}^{N\times d}\) and reorder them into a coarse spatial grid of size \((H/P)\times(W/P)\), restoring the original patch layout.  These grid features are then upsampled via a learnable transposed convolution layer to the full resolution \(H\times W\), producing the dense global feature map \(\mathbf{E}_{\mathrm{global}}\in\mathbb{R}^{H\times W\times d}\).  This process preserves the positional and contextual information captured by the ViT while generating a per‐pixel feature representation for downstream fusion.  

\paragraph{\textbf{Local detail via Convolutional Network.}}  
Concurrently, \(I\) is processed by three convolutional blocks, consisting of  \(M\) convolutional layers and instance normalization, with skip connections to capture fine‐scale texture and edge information.  Denoting \(\mathbf{E}_{\mathrm{local}}^{(0)}=I\), each layer computes,
\[
\mathbf{E}_{\mathrm{local}}^{(i)}
= \mathrm{Swish}\bigl(\mathrm{Conv}\bigl(\mathbf{E}_{\mathrm{local}}^{(i-1)};W_{\mathrm{conv}}^{(i)},b_{\mathrm{conv}}^{(i)}\bigr)\bigr),
\quad i=1,\dots,M,
\]
where \(\mathrm{Swish}(x)=x\,\sigma(\beta x)\) enhances low‐intensity details.  The output \(\mathbf{E}_{\mathrm{local}}^{(M)}\in\mathbb{R}^{H\times W\times d}\) emphasizes localized features.

\paragraph{\textbf{Fusion and mapping.}}  
To integrate global and local information, we simply add global feature map $\mathbf{E}_{global}$ and local feature map $\mathbf{E}_{local}$ as follows:
\[
\mathbf{E}_{\mathrm{sum}}(i,j)
= \mathbf{E}_{\mathrm{global}}(i,j) + \mathbf{E}_{\mathrm{local}}^{(M)}(i,j),
\]
then we map the combined hybrid features $\mathbf{E}_{sum} \in \mathbb{R}^{H \times W \times d}$ to the desired  feature space $\mathbf{E}_{img} \in \mathbb{R}^{H \times W \times \kappa}$ by $3 \times 3$ convolution (stride 1) followed by the Swish activation function. In compact form, we write  $\mathbf{E}_{img} = \Xi (\mathbf{E}_{sum})$, where  $\Xi : \mathbb{R}^{H \times W \times d} \rightarrow \mathbb{R}^{H \times W \times \kappa}$ is the convolution-plus-Swish mapping.

By combining the ViT’s hierarchical encoder–decoder with skip connections for global context and a convolutional branch for local precision, \(\mathcal{F}\) yields richly informative, spatially coherent embeddings that drive accurate 3D density prediction.  The detailed configurations of this hybrid ViT-CNN features extraction module are given in the table \ref{tab:model_config}.


\subsection{NeRF MLP and Rendering}

Unlike the original NeRF architecture \cite{mildenhall_nerf_2022}, our implementation of the density approximator \(\mathcal{M}\) introduces two key modifications adapted to our single-view 3D reconstruction task. First, we reduced the width of the hidden layer from 256 (as used in the original NeRF) to \(f=128\) to improve computational efficiency without sacrificing representational capacity. Second, while NeRF predicts both density and view-dependent RGB color, our model exclusively focuses on estimating the scalar density \(\widetilde{\rho}(r,s)\), simplifying the learning objective and making it more suitable for single-modality datasets such as PXs.

The density approximator \(\mathcal{M}\) is a fully‐connected network of depth \(\widetilde{d} =8\) and hidden width \(f=128\), which transforms each fused feature vector \(\mathbf{f}_{\mathrm{fused}}(r,s)\in\mathbb{R}^f\) into a scalar density \(\widetilde{\rho}(r,s)\).  We denote the layer activations by \(\mathbf{u}^{(\tilde{\ell})}\in\mathbb{R}^{f}\) for \(\tilde{\ell}=0,\dots,\widetilde{d}\), with $\mathbf{u}^{(0)} \;=\;\mathbf{f}_{\mathrm{fused}}(r,s)$.
Each hidden layer applies a linear map followed by the Swish activation \(\phi(x)=x\,\sigma(\beta x)\), chosen for its smooth gradient profile \cite{swi}. Mathematically it is expressed as:
\begin{equation}\label{eq:mlp-layer}
\mathbf{u}^{(\tilde{\ell}+1)}
= \phi\!\bigl(W^{(\tilde{\ell})}\,\mathbf{u}^{(\tilde{\ell})} + b^{(\tilde{\ell})}\bigr), 
\quad
\tilde{\ell}=0,\dots,\widetilde{d}-1.
\end{equation}
Here, each \(W^{(\tilde{\ell})}\in\mathbb{R}^{f\times f}\) learns to mix the \(f\) input features, and the bias \(b^{(\tilde{\ell})}\in\mathbb{R}^f\) shifts the activation to capture subtle density variations.

To preserve the original fused features and mitigate vanishing gradients, we insert a skip connection at layer \(\tilde{\ell}=4\).  We concatenate \(\mathbf{u}^{(4)}\) with the initial input \(\mathbf{u}^{(0)}\in\mathbb{R}^f\), forming a \(2f\)-dimensional vector $\mathbf{v}^{(4)} 
= \bigl[\mathbf{u}^{(4)};\,\mathbf{u}^{(0)}\bigr]\in\mathbb{R}^{2f},$
and compute
\begin{equation}\label{eq:mlp-skip}
\mathbf{u}^{(5)}
= \phi\!\bigl(W^{(4)}\,\mathbf{v}^{(4)} + b^{(4)}\bigr), 
\quad
W^{(4)}\in\mathbb{R}^{f\times 2f},\, b^{(4)}\in\mathbb{R}^f.
\end{equation}
This skip connection ensures that low‐level signals are directly available in deeper layers, improving learning stability and feature reuse \cite{mildenhall_nerf_2022}.

Finally, the last hidden state \(\mathbf{u}^{(\widetilde{d})}\) is linearly projected to a scalar and passed through a sigmoid to constrain \(\widetilde{\rho}\in(0,255)\) as follows:
\begin{equation}\label{eq:mlp-out}
\widetilde{\rho}(r,s) 
= \sigma\!\bigl(w_{\mathrm{out}}^\top\,\mathbf{u}^{(\widetilde{d})} + b_{\mathrm{out}}\bigr),
\quad
w_{\mathrm{out}}\in\mathbb{R}^f,\,b_{\mathrm{out}}\in\mathbb{R}.
\end{equation}

In summary, we collectively define an 8‐layer MLP with mid‐network feature fusion and smooth nonlinearities, combining the rich contextual information from \(\mathbf{f}_{\mathrm{fused}}\) with stable gradient flow—crucial for accurate density estimation in our neural radiance field.

\subsection{Refinement Module : 3D U-Net}

The refinement network $\mathcal{U}$ improves the coarse volume  $\widetilde{\rho}\in\mathbb{R}^{C\times H\times W\times D}$
to produce $\rho\in\mathbb{R}^{C\times H\times W\times D}$
using a standard 3D U-Net with feature‐channel dimensions  $[64,128,256,512]$ \cite{zhou_u-net_2020}. Briefly, the encoder consists of four convolutional blocks with output channels $h_1=64,\;h_2=128,\;h_3=256,\;h_4=512,$ where each block applies two $3\times3\times3$ convolutions (with instance‐norm and Swish activation) followed by a stride-2 downsampling convolution.  The decoder mirrors this with four upsampling blocks: each uses a $2\times2\times2$ transposed-conv to double spatial resolution, concatenates the corresponding encoder feature via a skip connection, and applies two further $3\times3\times3$ convolutions (with instance‐norm and Swish).

A final $1\times1\times1$ convolution reduces the 64-channel map back to $C=1$, and a sigmoid activation maps voxel values into $(0,255)$. This lightweight U-Net architecture efficiently refines coarse NeRF densities into anatomically coherent CBCT volumes while preserving fine spatial detail.  

\subsection*{Loss functions}






We define a composite training loss $\mathcal{L}$ that is a weighted sum of three components:  a volumetric mean squared error term, a projection consistency term, and a perceptual feature-based term. Formally, for a given input PX $I$, the total loss is expressed as 

\begin{align}
    \mathcal{L} \;=\; \mathcal{L}_{MSE} \;+\; \lambda_{1}\,\mathcal{L}_{proj} \;+\; \lambda_{2}\,\mathcal{L}_{perc}\,, \label{tot_loss}
\end{align}

where $\lambda_{1},\lambda_{2} \ge 0$ are weighting coefficients that regulate the influence of the projection and perceptual losses, respectively. Each term $\mathcal{L}_{MSE}$, $\mathcal{L}_{proj}$, and $\mathcal{L}_{perc}$ encourages a different aspect of agreement between the predicted 3D volume and the ground truth CBCT volume, as detailed below.

\paragraph{\textbf{Volumetric MSE Loss ($\mathcal{L}_{MSE}$).}} This term measures the voxel-wise fidelity of the reconstructed volume. Let $\rho(I,\mathbf{P})$ denote the predicted density at voxel location $\mathbf{P}=(x,y,z)$ given input $I$, and let $\hat{\rho}(I,\mathbf{P})$ be the corresponding ground-truth density. We define the volumetric mean squared error as:

\begin{align}
    \mathcal{L}_{MSE} \;=\; \frac{1}{|\Omega|}\sum_{\mathbf{P}\in \Omega} \Bigl(\rho(I,\mathbf{P}) \;-\; \hat{\rho}(I,\mathbf{P})\Bigr)^2\,,
\end{align}

where $\Omega$ is the set of all voxel coordinates in the 3D volume (with $|\Omega|$ being the number of voxels). This loss term directly penalizes local errors at each voxel, ensuring that the reconstructed density at every point in the volume closely matches the ground truth.

\paragraph{\textbf{Projection Consistency Loss ($\mathcal{L}_{proj}$).}} This term enforces global structural consistency by comparing the model’s 3D output to the ground truth via their radiographic projections. For a given volume $V$, let the maximum intensity projection (MIP) along a specified anatomical plane $v \in \{\text{axial},\,\text{sagittal},\,\text{coronal}\}$ be denoted by $\Pi_v(V)$. For instance, the axial projection is defined as

\begin{align}
    \Pi_{\text{axial}}(V)(x,y)=\max_{z}\; V(x,y,z),
\end{align}

with analogous definitions for the sagittal and coronal projections. The projection loss is then computed as:
\begin{align}
    \mathcal{L}_{proj} \;=\; \sum_{v \in \left\{ \substack{\text{axial},\\[3pt] \text{sagittal},\\[3pt] \text{coronal}} \right\}} \;\bigl\|\,\Pi_v\bigl(\rho(I)\bigr) \;-\; \Pi_v\bigl(\hat{\rho}(I)\bigr)\bigr\|_{2}^{2}\,,
\end{align}

where $\|\cdot\|_{2}^{2}$ denotes the sum of squared differences over all pixels in the projection image. This term ensures that the 2D projections derived from the predicted 3D volume align closely with those of the ground truth, thereby preserving the global anatomical structure.

\paragraph{\textbf{Perceptual Feature Loss ($\mathcal{L}_{perc}$).}} While the projection loss operates on raw pixel intensities, the perceptual loss compares high-level feature representations to account for perceptual realism. Leveraging a pretrained VGG-16 network $\Theta$, let $\Theta_{\ell}(I)$ denote the activation at layer $\ell$ for an input image $I$. For each view $v$, we compute the perceptual loss by comparing the feature representations of the predicted MIP $\Pi_v\bigl(\rho(I)\bigr)$ and the ground truth MIP $\Pi_v\bigl(\hat{\rho}(I)\bigr)$. It is mathematically expressed as,


\begin{align}
    \mathcal{L}_{perc} & \;=\; \nonumber\\
& \sum_{v \in \left\{ \substack{\text{axial},\\[3pt] \text{sagittal},\\[3pt] \text{coronal}} \right\}} 
\sum_{\ell \in L} 
\Bigl\| \,\Theta_{\ell}\bigl(\Pi_v\bigl(\rho(I)\bigr)\bigr) \;-\; 
\Theta_{\ell}\bigl(\Pi_v\bigl(\hat{\rho}(I)\bigr)\bigr)\Bigr\|_{2}^{2}\,.
\end{align}

where $L$ is a chosen set of VGG-16 layers. This loss term measures the discrepancy between the predicted and true projection images in the feature space, ensuring that the reconstruction preserves subtle anatomical details and textures.

\section{Experiments} \label{sec:expt}
\subsection{Experimental Setup}

\subsubsection{Implementation details}
\paragraph{\textbf{Dataset collection.} }  
The study used CBCT clinical head scans $623$, comprising $408$ women and $215$ men with an average age of $50$ years and a standard deviation of $26$ years. These scans were obtained from the Chosun School of Dentistry in Gwangju, South Korea. Image acquisition was performed with the Carestream Health CS900 scanner and the Planmeca VISOG7 system. The images were saved as $16$-bit DICOM slices, with a resolution range between $0.25 \, mm$ and $0.5 \, mm$. All data collection adhered to patient consent regulations and met the ethical standards established by The Code of Ethics of the World Medical Association. Institutional Review Board (IRB) approval was obtained prior to data collection.

\paragraph{\textbf{Dataset preparation.}}  The dataset preparation is outlined in Fig. \ref{fig:wf_data_pre}. Following our recent paper \cite{Anu_2024}, initially,
all CBCT scans are resampled to a common pixel spacing
of $0.3 \, mm$. Subsequently, the CBCT data is preprocessed to
obtain a contour box plot enclosing the jaw region. This
forms the region of interest (ROI) for 3D reconstruction. The ROI is cropped
out from CBCT and then reshaped to a uniform dimension
of $128 \times 256 \times 256$ ($  H\, \times  W\,  \times \,  D $). Similar to Park et al. \cite{park_2024}, a 
panoramic acquisition geometry is modeled and the ray
tracing is performed along the defined trajectory to
produce synthetic PX images of shape $128 \times 256$. Unlike Park et al.\cite{park_2024}, we have used an elliptical trajectory and rays are considered tangent to this trajectory. The number of sampled pixels along the rays are limited to a horseshoe shaped focal region enclosing the jaw. More details about the dataset preparations are provided in Appendix \ref{app:dp}. The model training utilizes the reshaped ROIs in conjunction with the synthesized PX images. Out of $623$ CBCT scans, $600$ patient data was generated using this approach and the data was split in a ratio $8:1:1$ for training, validation and testing, respectively.

\paragraph{\textbf{Data pre-processing.}}\label{pre_process}  
In high-dimensional medical imaging tasks, such as those involving 3D CBCT  and 2D PX images, preprocessing plays a crucial role in preparing the data for machine learning models. Here, we outline a series of mathematical transformations applied to standardize and normalize the pixel intensity distributions, ensuring numerical stability and compatibility with learning algorithms.

Let $\mathcal{I}\in\mathbb{R}^{C\times H\times W\times D}$ denote a single CBCT volume ($C=1,\, H=128,\, W=256$).  To mitigate extreme values, we first clip all intensities to lie between the 1\textsuperscript{st} and 99.9\textsuperscript{th} percentiles of the empirical distribution, followed by Z-score normalization (subtracting the global mean and dividing by the standard deviation) to center the data and enforce unit variance.  Finally, we linearly rescale the normalized values into \([0,255]\) to match typical 8-bit input ranges.  This concise pipeline—clipping, log‐compression, standardization, and min–max rescaling—suppresses outliers, symmetrizes the histogram, and stabilizes the numerical range, yielding consistent, well‐conditioned inputs for our 3D reconstruction network. 


\paragraph{\textbf{Model training configuration .}}

\begin{table}[htbp!]
	\centering
	\begin{tabularx}{\linewidth}{@{}lX@{}}
		\toprule
		\textbf{Parameter} & \textbf{Value} \\
		\midrule
		\midrule
		Deep Learning Framework  & PyTorch \\
		\addlinespace
		GPU Configuration       & NVIDIA RTX A6000 \\
		\addlinespace
		Training Time           & $\sim$22 hours 50 mins \\
		\addlinespace
		Optimizer               & Adam ($\beta_{1}=0.9,\;\beta_{2}=0.999$) \\
		\addlinespace
		Initial Learning Rate   & $1\times10^{-3}$ \\
		\addlinespace
		LR Scheduler            & ReduceLROnPlateau (monitor val.\ loss, factor 0.5, patience 15 epochs, min LR $1\times10^{-5}$) \\
		\addlinespace
		Batch Size              & 1\\
		\addlinespace
		Number of Epochs        & 300 \\
		\addlinespace
		Early Stopping          & Enabled (based on validation loss) \\
		\bottomrule
	\end{tabularx}
		\caption{Summary of training configurations.}
	\label{tab:training_config}
\end{table}

The reconstruction network was implemented in PyTorch and trained on eight NVIDIA RTX A6000 GPUs, requiring roughly 22 hours 50 minutes of wall‐clock time. We optimize with Adam (learning rate $1\times10^{-3}$, $\beta_{1}=0.9$, $\beta_{2}=0.999$) optimizer. A ReduceLROnPlateau scheduler monitors the validation loss and, if no improvement is observed for 15 epochs, multiplies the learning rate by 0.5 down to a floor of $1\times10^{-5}$. Training is run for 300 epochs with batch size 1 per GPU, and early stopping is enabled based on validation performance. The detailed summary of training configurations are given in the table ~\ref{tab:training_config}. Additionally, the detailed summary of  model configurations are also given in the appendix \ref{app:model_conf}.

\subsubsection{Evaluation metrics}
To assess image reconstruction results, the 2D-to-3D reconstruction model was evaluated with the standard metrices as follows: 
\begin{itemize}
    \item \textbf{PSNR :} Peak Signal-to-Noise Ratio (PSNR) measures the quality of a reconstructed image by comparing it to the original, focusing on the ratio between the maximum possible signal power and the power of corrupting noise. It's commonly used to evaluate lossy compression techniques.

    \item \textbf{SSIM :} The Structural Similarity Index Measure (SSIM) assesses image quality by focusing on structural information, luminance, and contrast, aiming to align with human visual perception \cite{SSIM}.   It's a full-reference metric comparing a distorted image to a pristine one.

    \item \textbf{LPIPS :} This is the Learned Perceptual Image Patch Similarity (LPIPS) which compares the generated and target images based on perceptual similarity \cite{zhang_2018}. For LPIPS, we use feature outputs from the pre-trained models -- VGG-16, AlexNet and SqueezeNet.

\end{itemize}

\subsection{Qualitative   Results}

\begin{figure*}[htbp!]
    \centering
    \includegraphics[width=\textwidth]{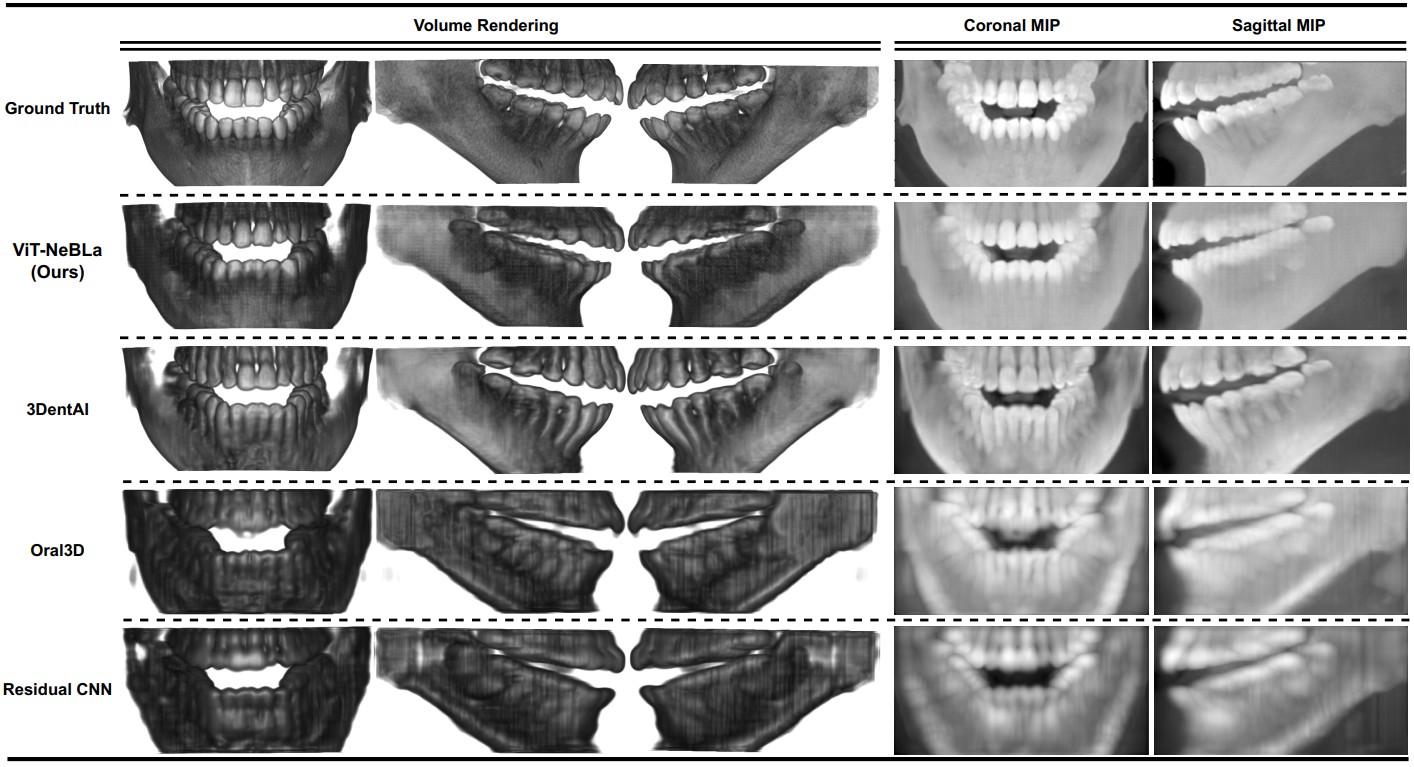}
    \caption{Qualitative comparison of reconstructed CBCT volumes. From top to bottom: ground truth; ViT-NeBLa (ours); 3DentAI \cite{3dentai}; Oral-3D auto-encoder \cite{song_2021}; and residual CNN \cite{henzler_2018}. Columns show (left) volume renderings, (middle) coronal MIPs, and (right) sagittal MIPs. Our ViT-NeBLa model yields the most accurate delineation of jaw anatomy—preserving cortical boundaries and fine internal structures—while suppressing artifacts and background noise compared to existing methods.}
    \label{fig:qc_1}
\end{figure*}

In Fig.~\ref{fig:qc_1} (See appendix \ref{app:visualization} for additional visualizations) we compare maximal intensity projections (MIPs) along the coronal and sagittal planes for four reconstruction approaches: (a) the autoencoder backbone of Oral-3D \cite{song_2021}, (b) a single-view residual CNN adapted from \cite{henzler_2018}, (c) the attention U-Net of 3DentAI \cite{3dentai}, and (d) our proposed method - ViT NeBLa.  All models were trained on the same synthetic PX inputs and under identical reconstruction protocols. Alongside the MIPs, we also present 3D volume renderings of each reconstructed density field, enabling a more holistic assessment of anatomical fidelity and volumetric consistency.

The residual CNN captures the coarse jaw shape but blurs key anatomical boundaries, yielding smooth yet indistinct mandibular edges.  Oral-3D and 3DentAI both improve edge clarity and overall silhouette accuracy, but occasionally exhibit spurious intensities near the condylar regions.  In contrast, our model consistently delineates the inferior border, gonial angle, and ramus with the greatest sharpness, while suppressing background artifacts.  This pronounced enhancement in both coronal/sagittal MIPs and volume-rendered views reflects our framework’s ability to integrate global context with local detail. Subtle ridges and foramina are clearly reconstructed, demonstrating a marked qualitative advantage over the baselines.



\subsection{Quantitative Results}

\begin{figure*}[htbp!]
	\centering
	\includegraphics[width=\linewidth]{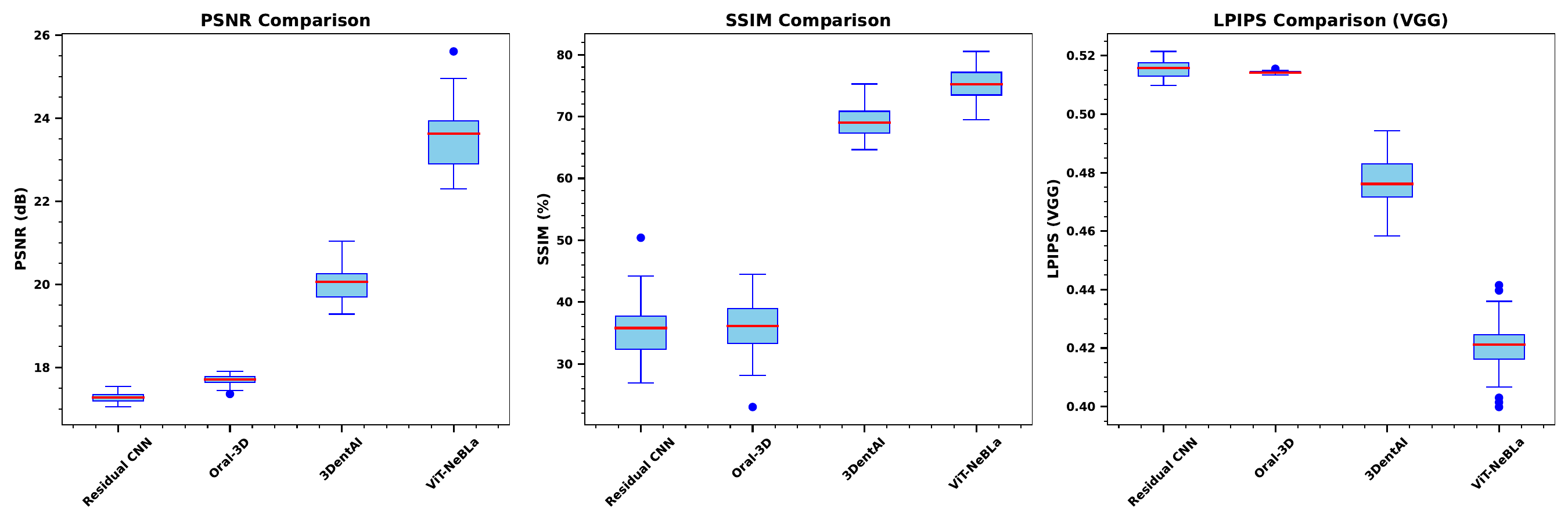}
	\caption{Boxplots of the PSNR (dB) ($\uparrow$), SSIM (\%)($\uparrow$), and LPIPS (VGG) ($\downarrow$) metrics for the four reconstruction methods evaluated in Table \ref{tab1}: Residual CNN \cite{henzler_2018}, Oral-3D \cite{song_2021}, 3DentAI \cite{3dentai}, and ViT-NeBLa (ours). Each subplot shows the distribution of  metric values based on the reported means  and standard deviation.}
	\label{fig:box_plot}
\end{figure*}

\begin{table*}[htbp!]
\centering
\renewcommand{\arraystretch}{1.2}%
\begin{tabular}{c|c|c|c|c}
\hline  \noalign{\smallskip}
 Method  &  Residual CNN \cite{henzler_2018} &  Oral-3D (auto-encoder) \cite{song_2021} & 3DentAI \cite{3dentai} & \textbf{ViT- NeBLa (Ours)}   \\
\hline \hline \noalign{\smallskip}
PSNR ($\uparrow$) & $17.30 \pm 0.13$ & $17.70 \pm 0.13$ & $20.05 \pm 0.40$ &  $\mathbf{23.4762 \pm 0.7823}$\\
SSIM(\%) ($\uparrow$) & $35.01 \pm 04.03$ & $36.05 \pm 04.20$ & $69.03 \pm 03.31$ &  $\mathbf{74.93 \pm 02.56}$\\
LPIPS (VGG) ($\downarrow$)& $0.5157 \pm 0.0028$ & $0.5143 \pm 0.0004$ & $0.4787 \pm 0.0082$ &  $\mathbf{0.4204 \pm 0.0093}$\\
\noalign{\smallskip}  \hline
\end{tabular}
\caption{Quantative comparison of the 3D oral reconstruction from single panoramic X-ray radiograph. The best results are highlighted in bold. The format is \textit{mean}$\pm$\textit{std} with $10$ repetitions of experiments.}\label{tab1}
\end{table*}

\begin{table*}[htbp!]
    \centering
    \renewcommand{\arraystretch}{1.2}%
    \begin{tabular}{c|c|c|c|c}
    \hline \noalign{\smallskip}
        LPIPS ($\downarrow$)& Residual CNN \cite{henzler_2018} & Oral-3D (auto-encoder) \cite{song_2021} & 3DentAI \cite{3dentai} & \textbf{ViT-NeBLa (Ours)}\\
        \hline \hline \noalign{\smallskip}
        VGG Net & $0.5157 \pm 0.0028$ & $0.5143 \pm 0.0004$ & $0.4787 \pm 0.0082$ & $\mathbf{0.4204 \pm 0.0093}$ \\
    Alex Net & $0.5804 \pm 0.0073$ & $0.5855 \pm 0.0018$ & $0.3553 \pm 0.0136$ & $\mathbf{0.3285 \pm 0.0143}$\\
        Squeeze Net & $0.3817 \pm 0.0055$ & $0.3804 \pm 0.0028$ & $0.3210 \pm 0.0112$ & $\mathbf{0.2581 \pm 0.0133}$\\
        \hline \noalign{\smallskip}
    \end{tabular}
    \caption{Comparision of perceptual similarity of the predicted CBCT volume from the models mentioned in the table. A lower value indicates higher perceptual similarity.}
    \label{tab:Lpips}
\end{table*}

In order to quantify reconstruction fidelity, we employ three complementary metrics - PSNR, SSIM, and LPIPS-which together gauge signal‐level accuracy, structural agreement, and perceptual similarity in learned feature spaces. As shown in Table \ref{tab1}, our ViT-NeBLa model achieves a PSNR of $23.4762 \pm 0.7823$ dB and an SSIM of $ 74.93 \pm 02.56$ \%, substantially outperforming the $20.05 \pm 0.40 \, \text{dB}/69.03 \pm 0.31$  \% of 3DentAI and the $ \approx 17.7 \, \text{dB}/36 $ \% of both the Oral-3D auto-encoder and the residual-CNN baselines. In perceptual terms, our method attains the lowest LPIPS (VGG) score of $0.4204 \pm 0.0093$ , indicating that its reconstructions most closely resemble ground truth when judged by deep network features. Table \ref{tab:Lpips} further corroborates this advantage across three different LPIPS backbones: compared to 3DentAI’s AlexNet LPIPS of $0.3553 \pm 0.0136$ , we reduce it to $0.3285 \pm 0.0143$ , and on SqueezeNet features from $0.3210 \pm 0.0112$  down to $0.2581 \pm 0.0133$. These statistics—each averaged over ten independent trials—demonstrate that fusing global transformer features with an implicit volumetric representation yields consistently superior quantitative performance. Figure \ref{fig:box_plot} visualizes the full distributions of PSNR, SSIM, and LPIPS across all methods via boxplots, highlighting the robustness of our improvements.

\subsection{Ablation studies}

\begin{table*}[!htbp]
    \centering
    \renewcommand{\arraystretch}{1.2}%
    \begin{tabular}{c|c|c|c}
    \hline \noalign{\smallskip}
        Method &  PSNR (dB)($\uparrow$) & SSIM (\%) ($\uparrow$) & LPIPS(VGG) ($\downarrow$)\\
    \hline \hline \noalign{\smallskip} 
       MSE  & $23.1095 \pm 0.7478$ & $73.69 \pm 02.66$ & $0.4356 \pm 0.0080$\\
       MSE + Proj & $23.1141 \pm 0.7088$ & $73.90 \pm 02.76$ & $0.4223 \pm 0.0105$\\
       MSE + Proj + Perc & $\mathbf{23.4762 \pm 0.7823}$ & $\mathbf{74.93 \pm 02.56}$ & $\mathbf{0.4204 \pm 0.0093}$ \\
    \hline \noalign{\smallskip} 
    \end{tabular}
    \caption{Ablation study on loss-function combinations. Quantitative metrics—PSNR (dB), SSIM (\%), and LPIPS (VGG)—are reported ($mean \pm std$ over ten runs) for three training configurations: MSE only, MSE + Projection consistency (Proj), and MSE + Proj + Perceptual feature loss (Perc). Adding projection consistency yields small SSIM and LPIPS improvements, while the full combination (MSE + Proj + Perc) achieves the highest PSNR (23.48 dB), best SSIM (74.93\%), and lowest perceptual error (LPIPS = 0.4204), confirming that each loss term contributes complementary gains.
    }
    \label{tab:ablation}
\end{table*}

\begin{figure*}[htbp!]
	\centering
	\includegraphics[width=\linewidth]{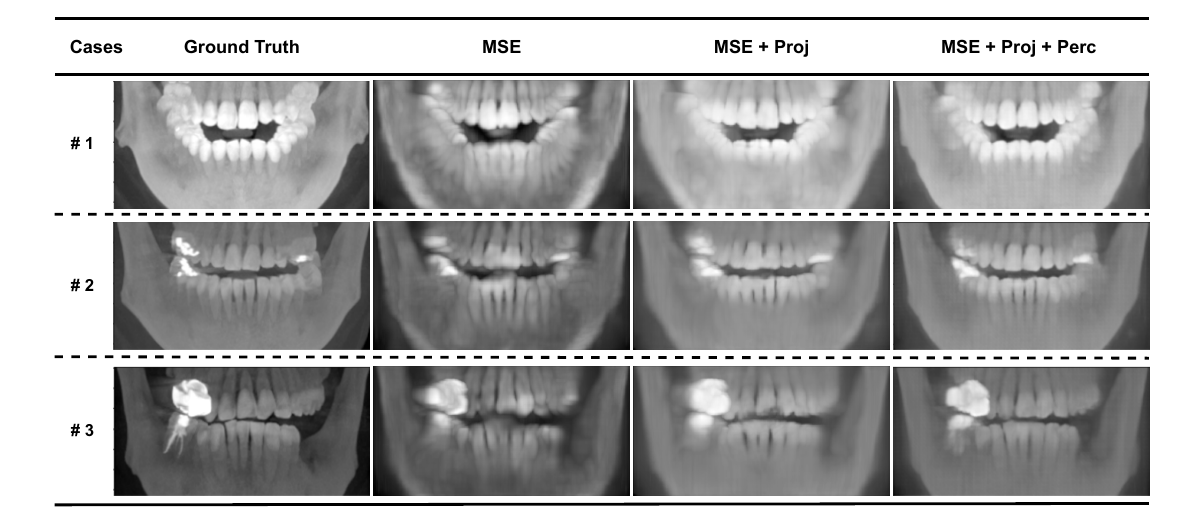}
	\caption{Ablation of loss components on 3D reconstruction quality. Each column shows coronal MIP outputs for three random patient cases (\# 1 – \# 3) under: (a) MSE loss only (PSNR = 23.11 dB, SSIM = 73.69 \%, LPIPS = 0.4356), (b) MSE + Projection consistency (PSNR = 23.11 dB, SSIM = 73.90 \%, LPIPS = 0.4223), and (c) full MSE + Projection + Perceptual loss (PSNR = 23.48 dB, SSIM = 74.93 \%, LPIPS = 0.4204), alongside the ground truth. Only the complete loss ensemble sharply delineates cortical boundaries and preserves fine trabecular patterns; the simpler loss configurations produce overly smooth volumes lacking subtle anatomical features.}
	\label{fig:ablation_study}
\end{figure*}

In order to  understand the individual contribution of each loss component to the quality of the predicted volumes, we performed an ablation study, the results of which are presented in Table~\ref{tab:ablation}. Starting from a pure voxel‐wise MSE loss, the model achieves $23.11$ dB PSNR, $73.69$ \% SSIM and an LPIPS of $0.4356$. Adding the projection consistency term yields a negligible PSNR change ($23.11$ dB) but improves SSIM to $73.90$ \% and reduces LPIPS to $0.4223$, indicating better adherence to radiographic projections. Crucially, when we incorporate the perceptual feature loss alongside MSE and projection, PSNR rises to $23.48$ dB, SSIM to $74.93$ \%, and LPIPS further drops to $0.4204$ -- our best quantitative result. More strikingly, Fig.~\ref{fig:ablation_study} shows that only the full “MSE + Proj + Perc” combination recovers sharp jaw borders, clear foramina, and coherent anatomical detail, whereas MSE alone or MSE + Proj still leave the volume looking overly smooth or lacking fine structure. This demonstrates that blending voxel‐level fidelity, projection‐level consistency, and perceptual realism produces the most visually and quantitatively compelling reconstructions.

\section{Limitations and future works}\label{sec:l&fw}


One limitation of the current approach lies in the discrepancy between synthetic and original PX images. While the synthesized PX images benefit from a dynamically adjustable focal trough that restricts the inclusion of irrelevant anatomical structures and ghost artifacts \cite{Anu_2024}, traditional PX images retain these unwanted features, resulting in higher noise and reduced model performance when used as input. This mismatch can lead to suboptimal predictions since the model is trained primarily on cleaner, synthetic data.

To address this, integrating an image-to-image translation module can prove beneficial. Such a module could transform the original PX images to more closely resemble their synthetic counterparts, thereby enhancing model robustness. While prior work by Park et al. \cite{park_2024} effectively utilized a cycle-consistent generative adversarial network (CycleGAN) for this purpose, future implementations could leverage more advanced techniques. Promising alternatives include diffusion models, ViTs  for GANs, StyleGAN3, and U-Net-based architectures with attention mechanisms. These models have demonstrated improved efficiency and performance in image-to-image translation tasks, potentially bridging the gap between real and synthetic PX inputs.

Another limitation is related to the stimulation of ray trajectories. While our current approach focuses on uniform sampling within the ROI defined by the focal trough, this method may limit prediction accuracy. Future work aims to implement dynamic sampling techniques to achieve more refined and accurate predictions.

Although our model is relatively lightweight compared to existing models for 3D reconstruction of oral cavities, which often require high-end GPUs, there is still room for improvement. Future efforts will focus on designing an even more lightweight, robust, and efficient model that can be trained on GPUs with lower memory capacity. This enhancement would not only reduce training and inference time but also make the model more accessible for broader applications.



\section{Conclusion}\label{sec:con}

In this work, we present a novel deep learning model called ViT-NeBLa, which combines vision transformers with the Neural Beer-Lambert model to achieve accurate 3D reconstruction directly from a single PX. We have shown that by restricting our NeRF‐based reconstruction to a predefined focal trough (ROI), we can dispense with any auxiliary density definition at each spatial sample—unlike prior work by Park et al. \cite{park_2024}. Inside this horseshoe‐shaped region, rays do not intersect, and all non‐ROI voxels are zeroed out. Consequently, we reduce per‐ray sampling from 200 points to just 96, cutting both memory and computation roughly in half without degrading reconstruction fidelity.

We further demonstrate that training on a more heterogeneous CBCT cohort—including cases with metallic implants and edentulous segments—yields a markedly more robust model than those evaluated on more uniform datasets. Despite this increased diversity, our ViT‐NeBLa consistently supersedes the state‐of‐the‐art in PSNR, SSIM, and LPIPS, indicating its capacity to generalize across a wide range of anatomical and artifact conditions.

Three key architectural innovations underpin this performance. First, we replace static Fourier encodings with a learnable multi‐resolution hash grid \cite{naf_2022}, achieving fine positional detail at constant memory cost. Second, our hybrid feature extractor pairs a ViT—capturing global context—with a lightweight CNN branch—capturing local texture—thereby outperforming conventional U‐Net backbones. Third, by eschewing any requirement for flattened CBCT volumes or explicit dental arch curves (unlike \cite{song_2021,liang_2021a,3dpx_2024,3dentai}), our model learns directly from  panoramic inputs and minimal annotation, broadening its clinical applicability.


Finally, while our implicit volumetric representation is highly memory‐efficient, it inherently obscures the explicit geometry favored in downstream workflows (e.g.\ mesh‐based surgical planning). As a next step, we plan to extend ViT‐NeBLa with an explicit 3D representation—such as Gaussian splatting \cite{gaussian_spl_2024}—to marry the best of both worlds-- the compactness of implicit fields and the interpretability of surface models. Together, these advances establish ViT‐NeBLa as a scalable, accurate, and clinically versatile system for recovering three‐dimensional oral anatomy from a single panoramic radiograph.

\section*{Acknowledgment}
We would like to thank Prof. Seong-Yong Moon from College of Dentistry in Chosun University  for facilitating the dataset.

\section*{Authors' Contributions}
\textbf{B.K. Parida}- Conceptualization of this study,
Methodology, Investigation, code implementation, experimentation, visualization, Writing - original draft, review
\& editing. \textbf{Anusree P.S. - } review \& editing. \textbf{A. Sen - } review \& editing. \textbf{W. You - } review \& editing,  funding acquisition, supervision.

\section*{Funding}
This work was supported by the Start-Up Growth Technology Development Project funded by the Ministry of Small and Medium Enterprises (SMEs) and Startups under Grant S3228660; in part by the Basic Science Research Program (NRF-2022R1F1A1075204) through the National Research Foundation of Korea (NRF) funded by the Ministry of Education, South Korea; and in part by the Regional Leading Company Innovation Promotion Program (S3455706) funded by the Ministry of Small and Medium Enterprises and Startups, South Korea.

\section*{Data Availability}
The datasets used in this study are not available publicly.

\section*{Additional Information}
 Requests for dataset should be request to W. You and the implemented code materials should be addressed to B. K. Parida. 

\appendix

\section{Dataset preparation: details \label{app:dp}}

\begin{figure*}[htbp!]
    \centering
    \includegraphics[width=0.9\linewidth]{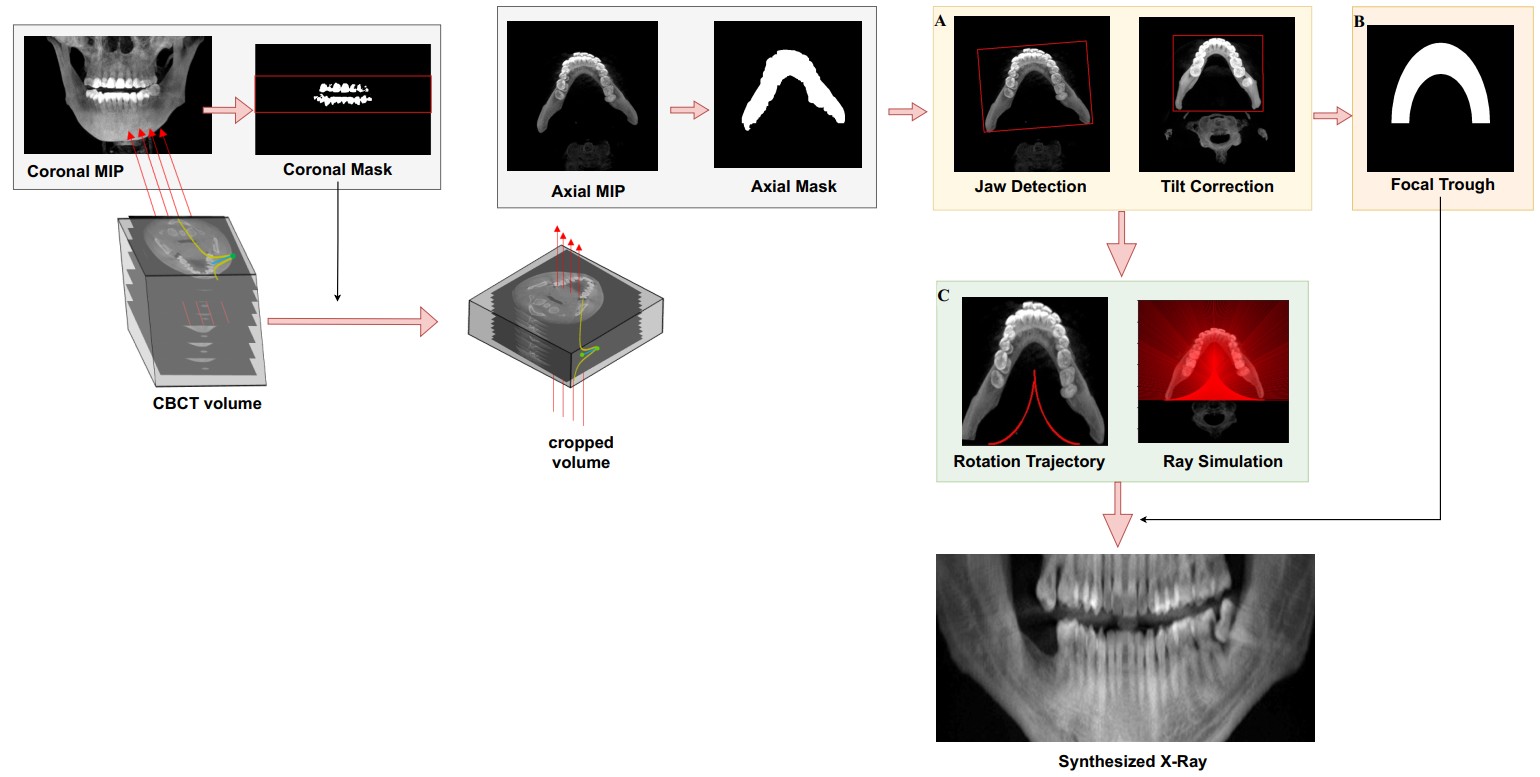}
    \caption{A complete workflow of the data preparation \cite{Anu_2024}: Coronal and Axial MIPs(Maximum Intensity Projection) images are generated to guide jaw detection. A. Perform Jaw contouring and horizontal tilt-correction. B. Define an elliptical focal trough region. C. Formulate dynamic rotation trajectory and Simulate pencil beams. D. Synthesized Panoramic X-ray. }
    \label{fig:wf_data_pre}
\end{figure*}

The proposed method \cite{Anu_2024}, illustrated in Fig. \ref{fig:wf_data_pre}, begins with detecting the jaw position from CBCT scans, which are complex compositions of tissues represented in grayscale influenced by various scanning factors. We applied rescaling and windowing techniques to highlight bone structures by setting a threshold above soft tissue values. Using Coronal  MIPs, we defined a  ROI encompassing the mandible and maxilla. An axial MIP refined this region, guiding contour extraction, with further morphological operations like opening and closing to handle artifacts from metal implants. To correct head tilt, contour plots were used, aligning the jaw with the sagittal plane to prevent asymmetrical magnification. Next, an elliptical focal trough was defined to capture the entire jaw, adapting in thickness to ensure clarity, especially around the incisors, minimizing ghost images. The scan simulation involved plotting an elliptical trajectory based on jaw contours, tracing rays tangentially to achieve a $180$-degree angular sweep with consistent sampling. Each ray, defined by Beer-Lambert’s law \cite{Ketcham_2014,Max_1995}, underwent attenuation integration over its length, discretized for practical computation. The final pixel intensity was calculated using transmittance adjustments, ensuring only relevant structures were rendered. This methodology, validated on a dataset of $600$ patient scans, demonstrated its robustness across diverse jaw shapes and artifacts, producing clear and clinically viable PXs.

\section{Detailed model configurations}  \label{app:model_conf}

\begin{table}[htbp!]
	\centering
	\begin{tabularx}{\linewidth}{@{}lX@{}}
		\toprule
		\textbf{Module} & \textbf{Configuration} \\
		\midrule
		\midrule
		Input image & $128\times256$ spatial, 1 channel \\
		\addlinespace
		\emph{CNN Extractor} & Each CNN block is 3 × \{Conv2d(in=1, out=256, 3×3, pad=1) + Swish($\beta$=1.2)\} followed by a skip connection. \\
		\addlinespace
		Patch embedding & Conv2d(in=1, out=256, kernel=16×16, stride=16) \\
		\quad\# patches & $(128/16)\times(256/16)=8\times16=128$ \\
		\addlinespace
		Position embedding & Learnable $\in\mathbb{R}^{1\times128\times256}$ \\
		\addlinespace
		\emph{Transformer Encoder} & 12 layers ($L$) of: \\
		& \quad• MultiheadAttention (embed\_dim=256, heads=8, dropout=0.1) \\
		& \quad• Feed-forward (256$\rightarrow$512$\rightarrow$256) with Swish, dropout=0.1 \\
		\addlinespace
		\emph{Transformer Decoder} & 12 layers mirrored to encoder (cross-add fusion) \\
		\bottomrule
	\end{tabularx}
	\caption{Detailed configuration of the hybrid ViT–CNN feature extractor.}
	\label{tab:model_config}
\end{table}

\begin{table}[htbp!]
	\centering
	\begin{tabularx}{\linewidth}{@{}lX@{}}
		\toprule
		\textbf{Parameter} & \textbf{Configuration} \\
		\midrule
		\midrule
		Input dimension  & 3 (3D coordinate $(x,y,z)$) \\
		\addlinespace
		Number of levels ($\xi$) & 16  \\
		\addlinespace
		Base resolution $(r_0)$  & 16 (lowest grid resolution) \\
		\addlinespace
		Maximum resolution ($r_{max}$) & 256 (highest grid resolution) \\
		\addlinespace
		Level resolutions ($r_{l}$) & $\{\min(16\cdot 2^i,\,256)\}_{i=0\ldots15}$ \\
		\addlinespace
		Features per level ($F$)  & 2 \\
		\addlinespace
		Hashmap size exponent  & 19 \\
		\addlinespace
		Prime numbers (hash seed) & [1,\,2654435761,\,805459861] \\
		\addlinespace
		Hash tables & 16 learnable tables of shape $[2^{19}\times 2]$, initialized $\mathcal{N}(0,0.01)$ \\
		\addlinespace
		Fast hash function & Bitwise‐XOR of scaled integer coordinates \\
		\addlinespace
		Concatenated output dim  & $3$ (input) + $16\times2$ (levels) = 35 \\
		\bottomrule
	\end{tabularx}
		\caption{Configuration of the learnable multi‐resolution hash positional  encoding module.}
	\label{tab:hash_encoding_config}
\end{table}


Our reconstruction framework comprises four sub‐modules namely (i) Hybrid ViT–CNN Feature Extractor, detailed in Table~\ref{tab:model_config}; (ii) Learnable Multi‐Resolution Hash Encoding, summarized in Table~\ref{tab:hash_encoding_config}; (iii) NeRF MLP Density Approximator, and (iv) 3D U‐Net Refinement Module.

\paragraph*{\textbf{NeRF MLP Density Approximator.}}
Each query point is first represented by a 128-dimensional vector. This vector is fed into an 8-layer MLP with hidden width 128, Swish activations, and a skip connection merging the input at layer 4. The final layer projects to a scalar density via a sigmoid, constraining $\rho\in[0,255]$. 

\paragraph*{\textbf{3D U-Net Refinement Module.}}The initial density volume $\tilde\rho\in\mathbb{R}^{C\times H\times W\times D}$ produced by the MLP is refined by a 3D U-Net with four encoder levels (each: two $3\times3\times3$ convs + InstanceNorm + Swish, followed by a stride-2 conv) and four decoder levels (each: $2\times2\times2$ transposed conv upsample, skip-connection concat, two $3\times3\times3$ convs + InstanceNorm + Swish). A final $1\times1\times1$ conv with sigmoid outputs the refined volume
 
To balance the multi‐term reconstruction loss in Eq.~\eqref{tot_loss}, we set 
\[
\lambda_{1} = \frac{1}{1.2}, 
\quad
\lambda_{2} = \frac{1}{25}\,. 
\]

\section{Additional Visualization Results}  \label{app:visualization}

\begin{figure*}[htbp!]
	\centering
	\includegraphics[width=\linewidth]{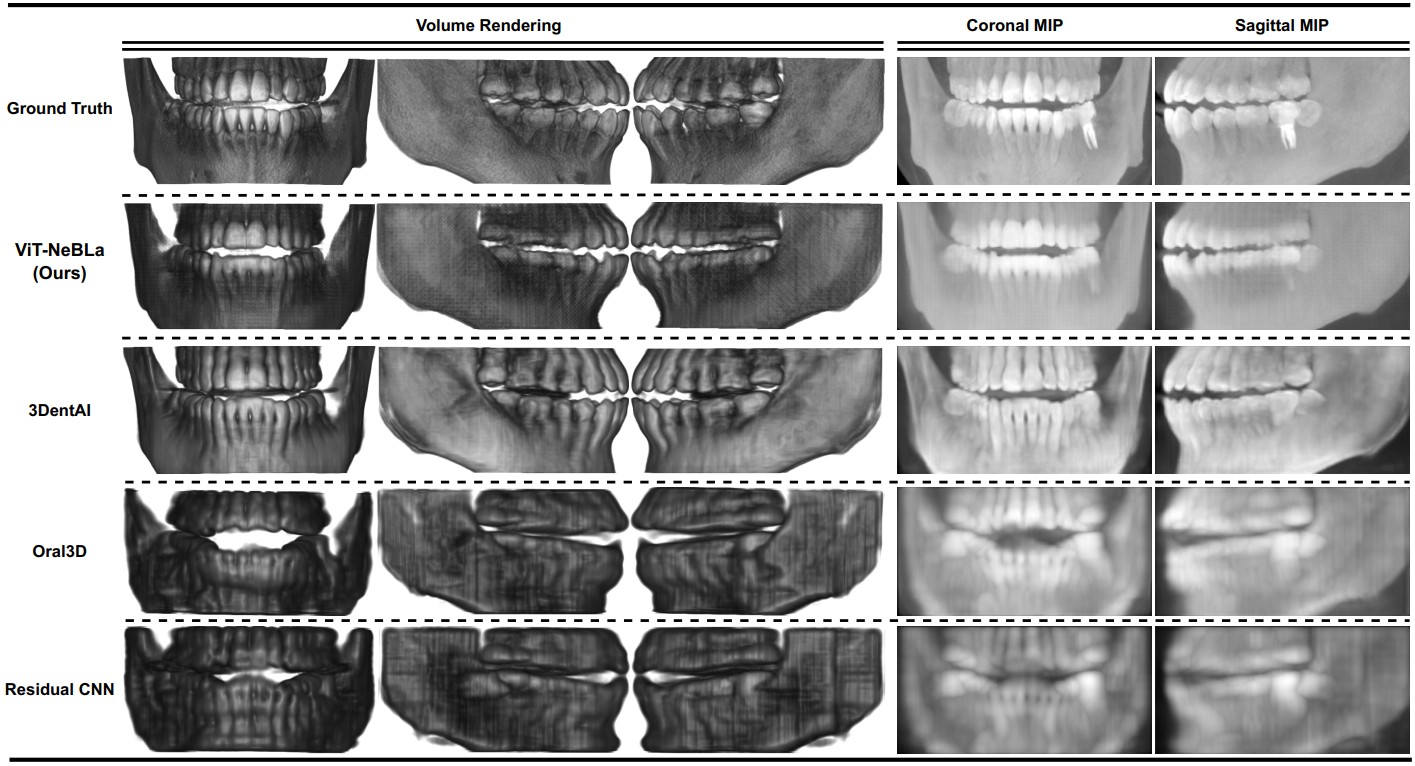}
	\caption{Qualitative comparison of reconstructed CBCT volumes. From top to bottom: ground truth; ViT-NeBLa (ours); 3DentAI \cite{3dentai}; Oral-3D auto-encoder \cite{song_2021}; and residual CNN \cite{henzler_2018}. Columns show (left) volume renderings, (middle) coronal MIPs, and (right) sagittal MIPs. 
	}
	\label{fig:qc_2}
\end{figure*}

\begin{figure*}[htbp!]
	\centering
	\includegraphics[width=\linewidth]{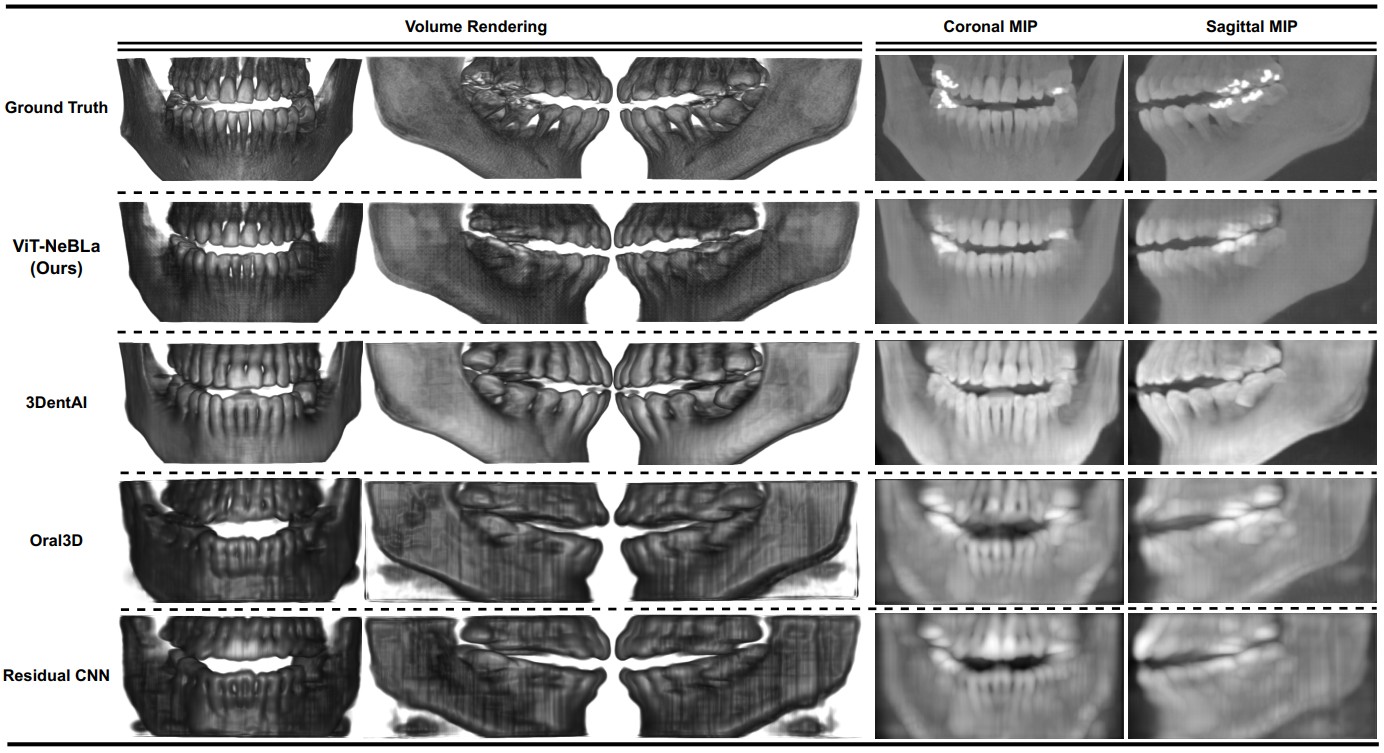}
	\caption{Qualitative comparison of reconstructed CBCT volumes. From top to bottom: ground truth; ViT-NeBLa (ours); 3DentAI \cite{3dentai}; Oral-3D auto-encoder \cite{song_2021}; and residual CNN \cite{henzler_2018}. Columns show (left) volume renderings, (middle) coronal MIPs, and (right) sagittal MIPs. 
	}
	\label{fig:qc_3}
\end{figure*}

\begin{figure*}[htbp!]
	\centering
	\includegraphics[width=\linewidth]{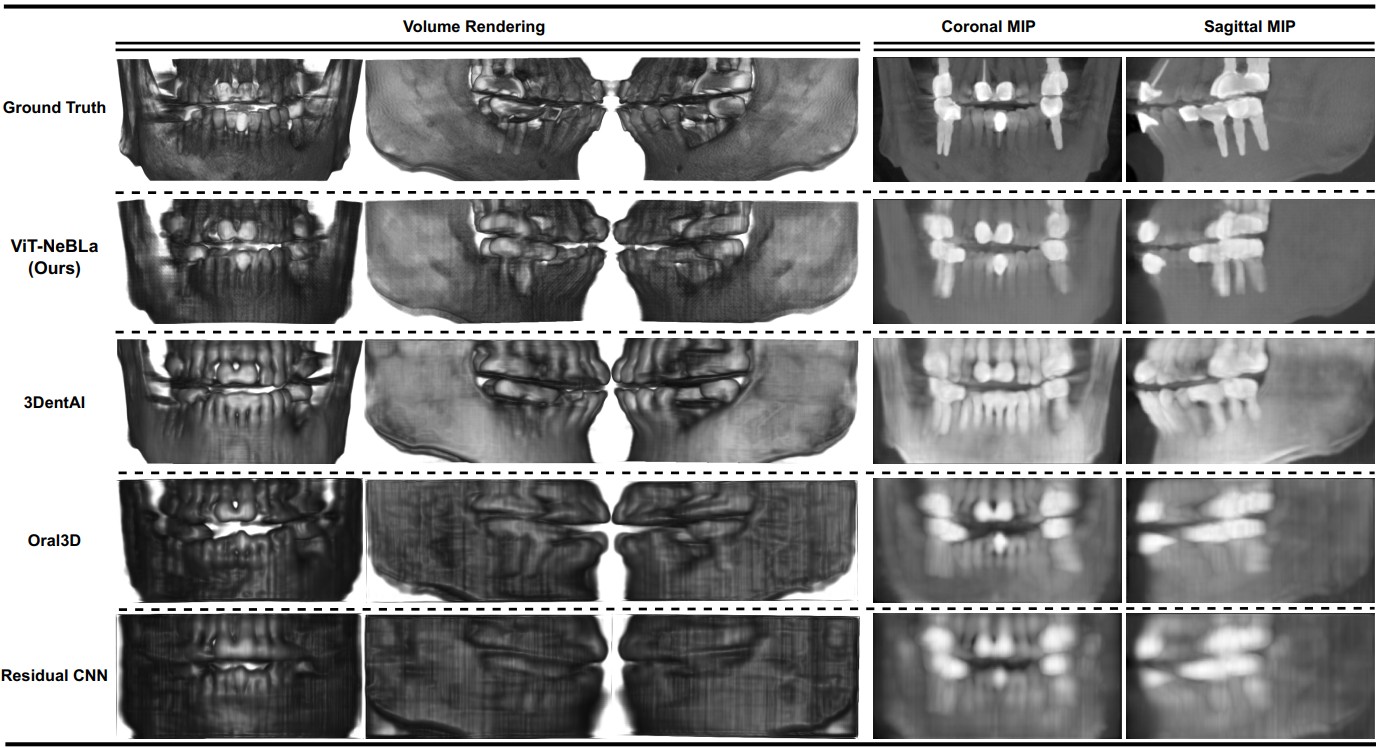}
	\caption{Qualitative comparison of reconstructed CBCT volumes. From top to bottom: ground truth; ViT-NeBLa (ours); 3DentAI \cite{3dentai}; Oral-3D auto-encoder \cite{song_2021}; and residual CNN \cite{henzler_2018}. Columns show (left) volume renderings, (middle) coronal MIPs, and (right) sagittal MIPs. 
	}
	\label{fig:qc_4}
\end{figure*}

\begin{figure*}[htbp!]
	\centering
	\includegraphics[width=\linewidth]{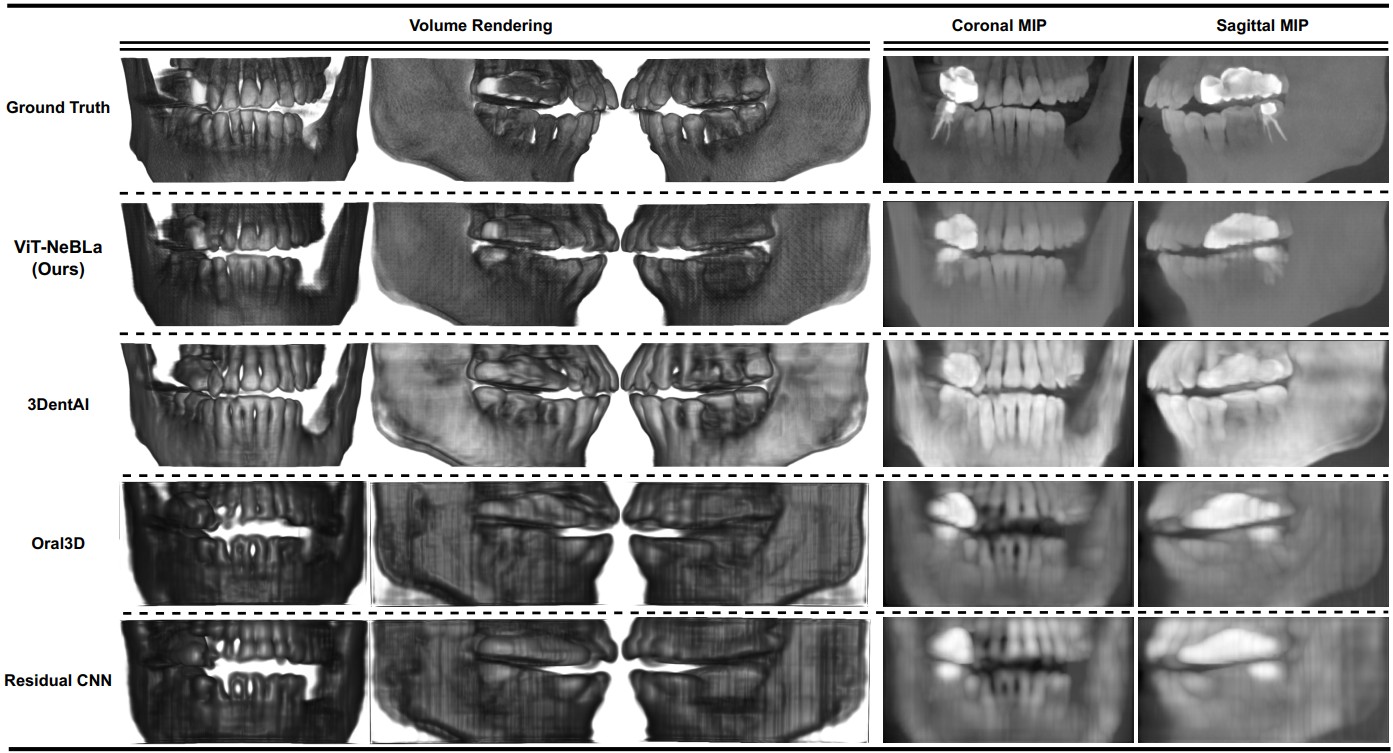}
	\caption{Qualitative comparison of reconstructed CBCT volumes. From top to bottom: ground truth; ViT-NeBLa (ours); 3DentAI \cite{3dentai}; Oral-3D auto-encoder \cite{song_2021}; and residual CNN \cite{henzler_2018}. Columns show (left) volume renderings, (middle) coronal MIPs, and (right) sagittal MIPs. 
	}
	\label{fig:qc_5}
\end{figure*}


In this section, we report additional qualitative reconstructions (Figs.~\ref{fig:qc_2}, \ref{fig:qc_3},\ref{fig:qc_4}, \& \ref{fig:qc_5}) to further assess our method’s robustness. Each figure presents both volume‐rendered views and MIPs for distinct PX inputs, illustrating how our pipeline consistently recovers detailed 3D anatomy—preserving cortical surfaces, fine trabecular patterns, and dental landmarks—while suppressing background artifacts across varied patient cases.

\clearpage  

\bibliographystyle{apsrev4-2}

\bibliography{References.bib} 

\end{document}